\documentclass[useAMS,usenatbib]{mnras}
\usepackage{journals_macros}
\usepackage{times}

\usepackage[english,english]{babel}
\usepackage{amsmath}
\usepackage{amssymb,amsfonts,textcomp}
\usepackage{array}
\usepackage{hhline}
\usepackage[usenames]{color}
\usepackage{xspace}
\usepackage{multirow}

\usepackage{graphicx}
\usepackage{natbib}
\usepackage{color}

\usepackage{times}
\usepackage{xspace}

\usepackage{enumitem}
\usepackage[compatibility=false]{caption}
\usepackage{subcaption}
\usepackage{tabulary}
\usepackage[para]{threeparttable}
\usepackage{array,booktabs,longtable,tabularx}
\newcolumntype{L}{>{\raggedright\arraybackslash}X}
\usepackage{ltablex}
\renewlist{tablenotes}{enumerate}{1}
\makeatletter
\setlist[tablenotes]{label=\tnote{\alph*},ref=\alph*,itemsep=\z@,topsep=\z@skip,partopsep=\z@skip,parsep=\z@,itemindent=\z@,labelindent=\tabcolsep,labelsep=.2em,leftmargin=*,align=left,before={\footnotesize}}
\makeatother

\def\gtrsim{\lower.5ex\hbox{$\; \buildrel > \over \sim \;$}}
\usepackage{graphicx}

\newcommand{\msun}{\mbox{$M_\odot$}}

\newcommand{\hagn}{\mbox{{\sc \small Horizon-AGN}}\xspace}
\newcommand{\hnoagn}{\mbox{{\sc \small Horizon-noAGN}}\xspace}

\newcommand{\ssfr}{\mbox{sSFR}\xspace}
\newcommand{\vsig}{\mbox{V/$\sigma$}\xspace}

\def \apj {{\it ApJ}}
\def \apjl {{\it ApJ Let.}}
\def \apjs {{\it ApJ Sup.}}
\def \pasj {{\it PASJ}}

\def \aap {{\it A\&A}}
\def \aj {{\it AJ}}
\def \mnras {{\it MNRAS}}
\def \nat {{\it Nature}}

\usepackage{hyperref}

\begin{document}

\author[K. Kraljic, C. Pichon, Y. Dubois, S. Codis et al.]{\parbox[t]{\textwidth}{K. Kraljic$^{1}$, C. Pichon$^{1,2,3}$, Y. Dubois$^{2}$, S. Codis$^{2}$, C. Cadiou$^{2}$, J. Devriendt$^{4,5}$, \\ M. Musso$^{6}$, C. Welker$^{7}$,
S. Arnouts$^{8}$, H. S. Hwang$^{9}$, C. Laigle$^{4}$, S. Peirani$^{2,10}$,\\
 A. Slyz$^{4}$,  M. Treyer$^{8}$ and D. Vibert$^{8}$}
\vspace*{6pt} \\
$^{1}$ Institute for Astronomy, University of Edinburgh, Royal Observatory, Blackford Hill, Edinburgh, EH9 3HJ, United Kingdom\\
$^{2}$ CNRS and Sorbonne Universit\'e, UMR 7095, Institut d'Astrophysique de Paris, 98 bis Boulevard Arago, F-75014 Paris, France\\
$^{3}$ School of Physics, Korea Institute for Advanced Study (KIAS), 85 Hoegiro, Dongdaemun-gu, Seoul, 02455, Republic of Korea\\
$^{4}$ Sub-department of Astrophysics, University of Oxford, Keble Road, Oxford, OX1 3RH, United Kingdom\\
$^{5}$ Observatoire de Lyon, UMR 5574, 9 avenue Charles Andr\'e, Saint Genis Laval 69561, France\\
$^{6}$ East African Institute for Fundamental Research (ICTP-EAIFR), KIST2 Building, Nyarugenge Campus, University of Rwanda, Kigali, Rwanda\\
$^{7}$ {International Centre for Radio Astronomy Research and  ASTRO 3D, University of Western Australia, 35 Stirling Highway, Crawley, WA 6009, Australia}\\
$^{8}$ {Aix Marseille Univ, CNRS, LAM, Laboratoire d'Astrophysique de Marseille, 38 Rue Fr\'ed\'eric Joliot Curie, 13013, Marseille, France} \\
$^{9}$ Quantum Universe Center, Korea Institute for Advanced Study, 85 Hoegiro, Dongdaemun-gu, Seoul 02455, Republic of Korea\\
$^{10}$ Laboratoire Lagrange, UMR7293, Universit\'e de Nice Sophia Antipolis, CNRS, Observatoire de la C\^ote d'Azur, 06300 Nice, France
}
\date{Submitted to MNRAS 2018 August 14}

\title[Galaxies in the saddle frame of the cosmic web]{Galaxies flowing in the oriented saddle frame of the cosmic web}

\maketitle

\begin{abstract}
{
The strikingly anisotropic large-scale distribution of matter made of an extended network of voids delimited by sheets, themselves segmented by filaments, within which matter flows towards compact 
nodes where they intersect, imprints its geometry on the dynamics of cosmic flows, ultimately shaping the distribution of galaxies and the redshift evolution of their properties.
The (filament-type) saddle points of this cosmic web provide a local frame in which to quantify the induced physical and morphological evolution of galaxies on large scales.
The  properties of virtual galaxies within the \hagn\,simulation are stacked in such a frame.
The iso-contours of the galactic number density, mass, specific star formation rate (sSFR), kinematics and age are  clearly aligned with the filament axis with steep gradients perpendicular to the filaments.
A comparison to a simulation without feedback from active galactic nuclei (AGN) illustrates its impact on quenching star formation of centrals away from the saddles.
The redshift evolution of the properties of galaxies and their age distribution are consistent with the geometry of the bulk flow within that frame.
They compare  well with expectations from constrained Gaussian random fields and the  scaling with the  mass of non-linearity, modulo the redshift dependent impact of feedback processes. 
Physical properties such as \ssfr and kinematics seem {\sl not} to depend {\sl only} on mean halo mass and density: the residuals trace the geometry of the saddle,
which could point to other environment-sensitive physical processes, such as spin advection, and AGN feedback at high mass. }
\end{abstract}

\begin{keywords}
galaxies: formation ---
galaxies: evolution ---
galaxies: interactions ---
galaxies: kinematics and dynamics ---
methods: numerical
\end{keywords}

\section{Introduction}
\label{section:introduction}

\
\begin{figure*}
\centering\includegraphics[width=0.85\textwidth]{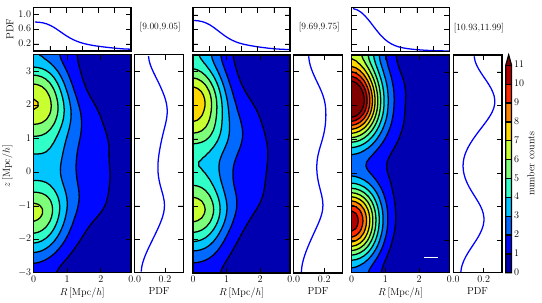}
\caption{The galaxy number counts in low ({\sl left}), intermediate ({\sl middle}) and high ({\sl right}) stellar mass bins (see text for definition), as labeled (in square brackets), at redshift zero in the frame of the closest saddle. The vertical axis corresponds to the distance from the saddle along the skeleton, while the horizontal axis corresponds to the transverse direction to the skeleton. The upward direction is defined as the direction of the node with the highest density.
The white horizontal line represents the smoothing length used in the analysis. The sub-panels on the top and on the right of each panel show the marginalised 1D distributions of $R$ and $z$, respectively. Note that the behaviour of the gradient of the number density of galaxies changes with stellar mass, noticeably in high mass bin. 
As expected, high mass galaxies are more tightly clustered near the filament axis and near nodes ({\sl right panel}) compared to their lower mass counterparts ({\sl middle }and {\sl left panels}).
 }
\label{fig:2D_curv_761_nb_allM_mass}
\end{figure*}

Galaxies form and evolve within a complex network, the so-called cosmic web \citep{Bond1996}, made of filaments embedded in sheet-like walls,  surrounded by large voids and intersecting at clusters of galaxies \citep{Joeveer1978}. 
Do the properties of galaxies, such as their morphology, retain a memory of these large-scale cosmic flows from which they emerge?
The importance of interactions with the larger scale environment in driving their evolution has indeed recently emerged as central tenet of galaxy formation theory.
Galactic masses are highly dependent on their large-scale surrounding, as elegantly explained by the theory of biased clustering \citep{Kaiser1984,Efstathiou1988}, such that high mass objects preferentially form in over-dense environment near nodes \citep{BondMyers1996,Pogosyan1996}.
Conversely, what are the signatures of this environment away from the nodes of the cosmic web?

While galaxies grow in mass when forming stars from intense gas inflows at high-redshift, they also acquire spin through tidal torques and mergers biased by these anisotropic larger scales
\citep[e.g.][for dark matter, and \citealt{Pichon2011,Dubois2014,Welker2014}, in hydrodynamical simulations]{Aubert2004,peirani04,Navarro2004,AragonCalvo2007,Codis2012,Libeskind2012,Stewart2013,Trowland2013,AragonCalvo2014}.
This should in turn have a significant impact on galaxy properties including morphology, colour and star formation history of galaxies.

As a filament is formally the field line that joins two maxima of the density field through a filament-type saddle point\footnote{where the gradient of the density field is null and the density Hessian has two negative eigenvalues} \citep{Pogosyan2009},
studying the  expected properties of galaxies \textit{in the vicinity of  filament-type saddle points}  is a sensible choice.
Indeed, Tidal Torque Theory \citep{Peebles1969,schaefer08} was recently revisited \citep{Codis2015a} in the context of such anisotropic environments, biased by the presence of a filament within a wall, which is most efficiently represented by this point process of filament-type saddles\footnote{The constrained misalignment between the tidal and the inertia tensors in the vicinity of  filament-type saddles simply explains the distribution of spin directions and its mass dependent flip.}. It predicts the alignment of the angular momentum distribution of the forming galaxies with the filament's direction, and perpendicular orientation for massive population.
Since spin  plays an important role in the physical and morphological properties of galaxies, a signature is also expected in the properties of galaxies as a function of the longitudinal and transverse distance to this saddle.

Most of the previous theoretical work on the impact of the anisotropy of the environment on galactic assembly history focused on dark matter haloes.
In the emerging picture of halo assembly history, at a given mass, haloes that are sufficiently far from the potential wells of other haloes can grow by accretion from their neighbourhood, leading to a  correlation between the accretion rate of haloes and the density of their environment \citep[e.g.][]{Zentner2007}. Haloes that are close to more massive structures are on the other hand expected to be stalled and their growth may stop earlier, as their mass inflow is dynamically quenched by anisotropic tides generated in their vicinity \citep[e.g.][]{Dalal2008,Hahn2009,Ludlow2014,Borzyszkowski2017,Paranjape2018a}.
Individual properties of dark matter haloes, such as their mass, formation time or accretion, are thus expected to be affected by the exact position of haloes within the large-scale anisotropic cosmic web \citep[e.g.][]{Lazeyras2017}.
Such  expectations are complementary  to the recent work of \cite{Musso2018}
whose analytical prediction of the mass, accretion rate and formation time of dark matter haloes near proto-filaments (identified as saddle points of the gravitational potential field) confirms that the anisotropy of the cosmic web is a significant ingredient to describe jointly the dynamics and physics of haloes. Their model predicts that at {\sl fixed} mass, mass accretion rate and formation time of haloes also vary with orientation and distance from the saddle.

Theoretical predictions on the impact of the anisotropic tides of the cosmic web on the specific properties of {\sl galaxies} embedded in those haloes are hampered by the complexity of baryonic processes and the lack of knowledge of detailed physics driving them.
Some attempts were recently made by \cite{Alam2018} and \cite{Paranjape2018b} which  compared the observed clustering and quenching properties of galaxies in the Sloan Digital Sky Survey with corresponding measurements in mock galaxy catalogues. These studies focused on whether the cosmic web leaves an imprint on the galaxy clustering beyond the effects of halo mass, by constructing mock catalogues using a halo occupation distribution in such a way that dependencies of galaxy properties on the tidal anisotropy and isotropic overdensity are driven by the underlying halo mass function across the cosmic web alone. As such prescription qualitatively reproduces the main observed trends, and quantitatively matches many of the observed results, they concluded that any additional direct effect of the large-scale tidal field on galaxy formation must be extremely weak.

In this work, the adopted approach is different in that it focuses directly on galaxies, their physical properties and redshift evolution as measured in the large-scale cosmological hydrodynamical simulation Horizon-AGN~\citep{Dubois2014, Dubois2016}. The main purpose of this paper is to show how the 3D distribution of the physical properties of these synthetic galaxies reflects the (tidal) impact of the cosmic web on the assembly history of galaxies.
 It is partly motivated by recent studies carried in the VIPERS, GAMA and COSMOS surveys
\citep{Malavasi2017,Kraljic2018,Laigle2018}
which showed that the colour and specific star formation rate  of galaxies are sensitive to their proximity to the cosmic web at fixed stellar mass
and local density. This paper focuses specifically on the distribution of the galaxy properties stacked in the oriented frame of the filament on large ($\sim$ Mpc) scales.
The natural choice of  frame for  stacking is defined by filament-type saddle points connecting two nodes by one filament (in contrast to nodes which are typically places where the connectivity of filaments is higher).

This paper is organised as follows. Section~\ref{section:methods} shortly describes the simulation and the detection of filaments within.
Section~\ref{section:3Dgalaxies} presents the galactic maps near the saddle, focusing first on
the transverse and longitudinal (azimuthally averaged) maps and then their 3D counterparts, while Section~\ref{section:redshift} shows their redshift evolution. Section~\ref{sec:theory} relates our finding to the properties  of weakly non-Gaussian random fields near saddles.
Some observational implications of our work together with the comparison with theoretical predictions are discussed in Section~\ref{section:discussion}.
Section~\ref{section:conclusion} wraps up. \\
Appendix~\ref{sec:validation} explores the robustness of our finding
w.r.t. smoothing and choice of filament tracer, Appendix~\ref{sec:filaments} discusses the redshift evolution of the geometry of filaments, 
Appendix~\ref{sec:azimuthal} presents complementary 2D maps,  Appendix~\ref{sec:AGN_noAGN_ssfr} quantifies the position-in-the-saddle frame efficiency of AGN feedback.
Appendix~\ref{section:theory} sketches the derivation of the theoretical results presented in the main text.
Appendix~\ref{sec:vel} presents the geometry of the bulk galactic velocity flow in the frame of the saddle.
Finally,  Appendix~\ref{sec:stat} motivates statistically the mediation of 
 mass and density maps over tides.
Throughout this paper, by $\log$, we refer to the 10-based logarithm and we loosely use $\log\! M$ as a short term for $\log(M/M_{\sun})$ and $\log\! \rho$ for $\log(\rho/M_{\sun} h^{-2} \rm{Mpc}^3)$.

\section{Numerical methods}
\label{section:methods}

Let us briefly review the main numerical tools used in this work to study the properties of virtual galaxies within the frame of the saddle.

\begin{figure}
\centering\includegraphics[width=0.85\columnwidth]{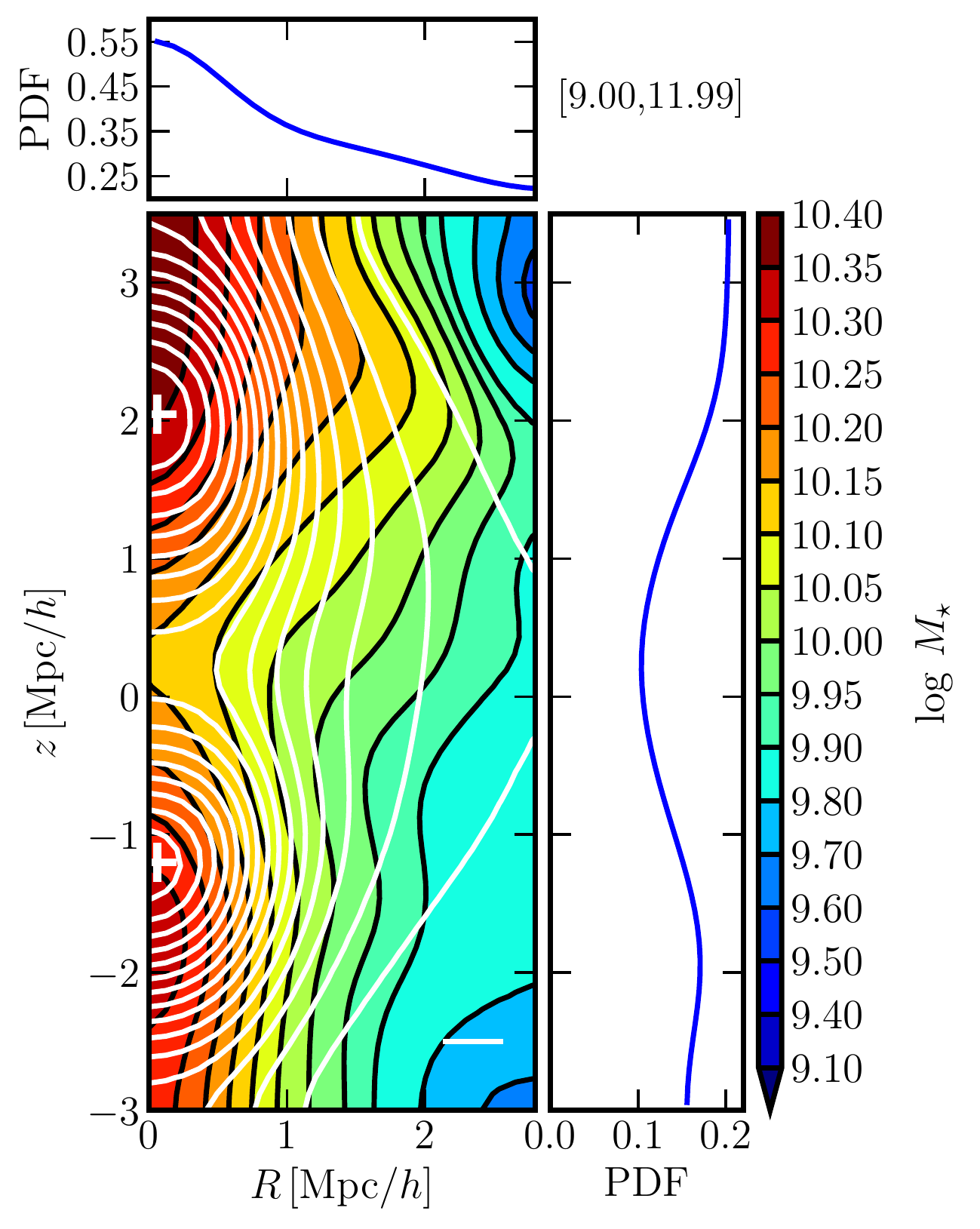}
\caption{Mean stellar mass in the frame of the closest saddle for the entire galaxy population with masses in the range $10^{9}$ to $10^{12} \msun$ at redshift zero. The white curves correspond to the contours of the galaxy number counts, while the white crosses represent the peaks in galactic number density on axis. More massive galaxies are further away from the saddle (resp. closer to the saddle) than the low mass population in longitudinal (resp. transverse) direction.
}
\label{fig:2D_curv_761_nb_allM_mass2}
\end{figure}


\subsection{Hydrodynamical simulation}

The details of the  \hagn\,simulation\footnote{see \href{http://www.horizon-simulation.org}{http://www.horizon-simulation.org}} can be found in~\cite{Dubois2014}, here, only brief description is provided. The simulation is performed with the Adaptive Mesh Refinement code {\sc ramses}~\citep{Teyssier2002} using a box size of $100\, h^{-1}\, \rm Mpc$ and adopting a $\Lambda$CDM cosmology with total matter density $\Omega_{\rm  m}=0.272$, dark energy density $\Omega_\Lambda=0.728$, baryon density $\Omega_{\rm  b}=0.045$, amplitude of the matter power spectrum $\sigma_8=0.81$, Hubble constant $H_0=70.4 \, \rm km\,s^{-1}\,Mpc^{-1}$, and $n_s=0.967$ compatible with the WMAP-7 data~\citep{Komatsu2011}. 
The total volume contains $1024^3$ dark matter (DM) particles, corresponding to a dark matter mass resolution of $M_{\rm  DM, res}=8\times10^7 \, \rm M_\odot$. The initial gas resolution is $M_{\rm gas,res}=1\times 10^7 \, \rm M_\odot$. 
The refinement of the initially coarse $1024^3$ grid down to $\Delta x=1$ proper kpc is triggered in a quasi-Lagrangian manner: if the total baryonic mass reaches 8 times the initial dark matter mass resolution, or the number of dark matter particles becomes greater than 8 in a cell, resulting in a typical number of $7\times 10^9$ gas resolution elements (leaf cells) at redshift zero.

The gas heating from a uniform ultraviolet background that takes place after redshift $z_{\rm  reion} = 10$ is modelled following~\cite{Haardt1996}. 
Gas is allowed to cool down to $10^4\, \rm K$ through H and He collisions with a contribution from metals~\citep{Sutherland1993}.
Star formation follows a Schmidt relation in regions of gas number density above $n_0=0.1\, \rm H\, cm^{-3}$: $\dot \rho_*= \epsilon_* {\rho_{\rm g} / t_{\rm  ff}}$,  where $\dot \rho_*$ is the star formation rate mass density, $\rho_{\rm g}$ the gas mass density, $\epsilon_*=0.02$ the constant star formation efficiency, and $t_{\rm  ff}$ the local free-fall time of the gas.
Feedback from stellar winds, supernovae type Ia and type II are included into the simulation with mass, energy and metal release \citep[see][for further details]{Kaviraj16}.

The \hagn simulation includes the formation of black holes (BHs) that can  grow by gas accretion at a Bondi-capped-at-Eddington rate and coalesce when they form a tight enough binary. Energy of BHs can be released in a heating or jet mode (respectively ``quasar'' and ``radio'' mode) when the accretion rate is respectively above and below one per cent of Eddington, with efficiencies tuned to match the BH-galaxy scaling relations at redshift zero~\citep[see][for further details]{Dubois2012}.

In order to assess the impact of Active Galactic Nuclei (AGN) feedback on galaxy properties in the frame of saddle, this analysis also relies on the \hnoagn simulation, which was performed with identical initial conditions and sub-grid modelling, but without BH formation, thus without AGN feedback \citep{Dubois2016, Peirani17}.

\begin{figure*}
\centering
\includegraphics[width=0.45\textwidth]{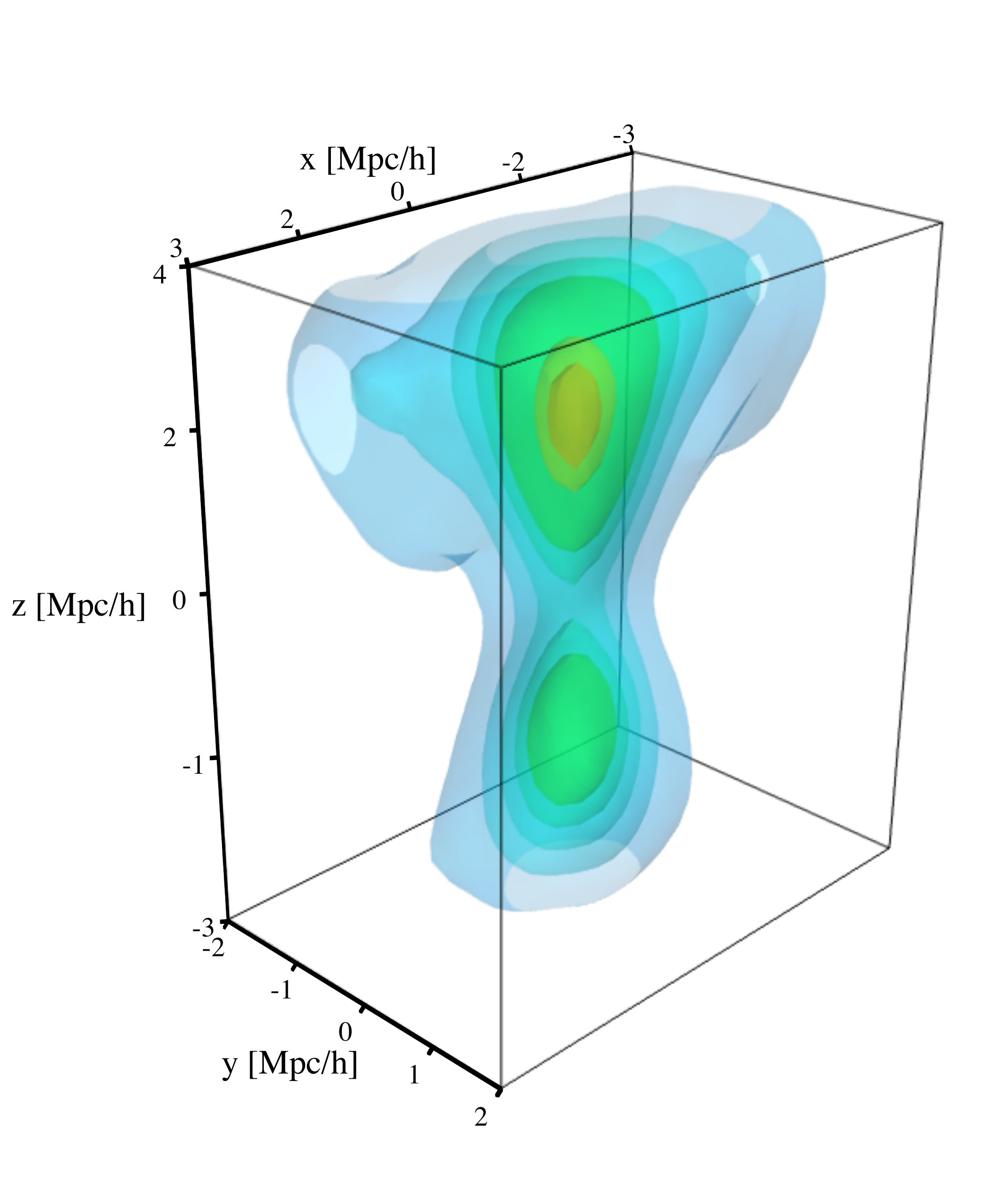}
\includegraphics[width=0.45\textwidth]{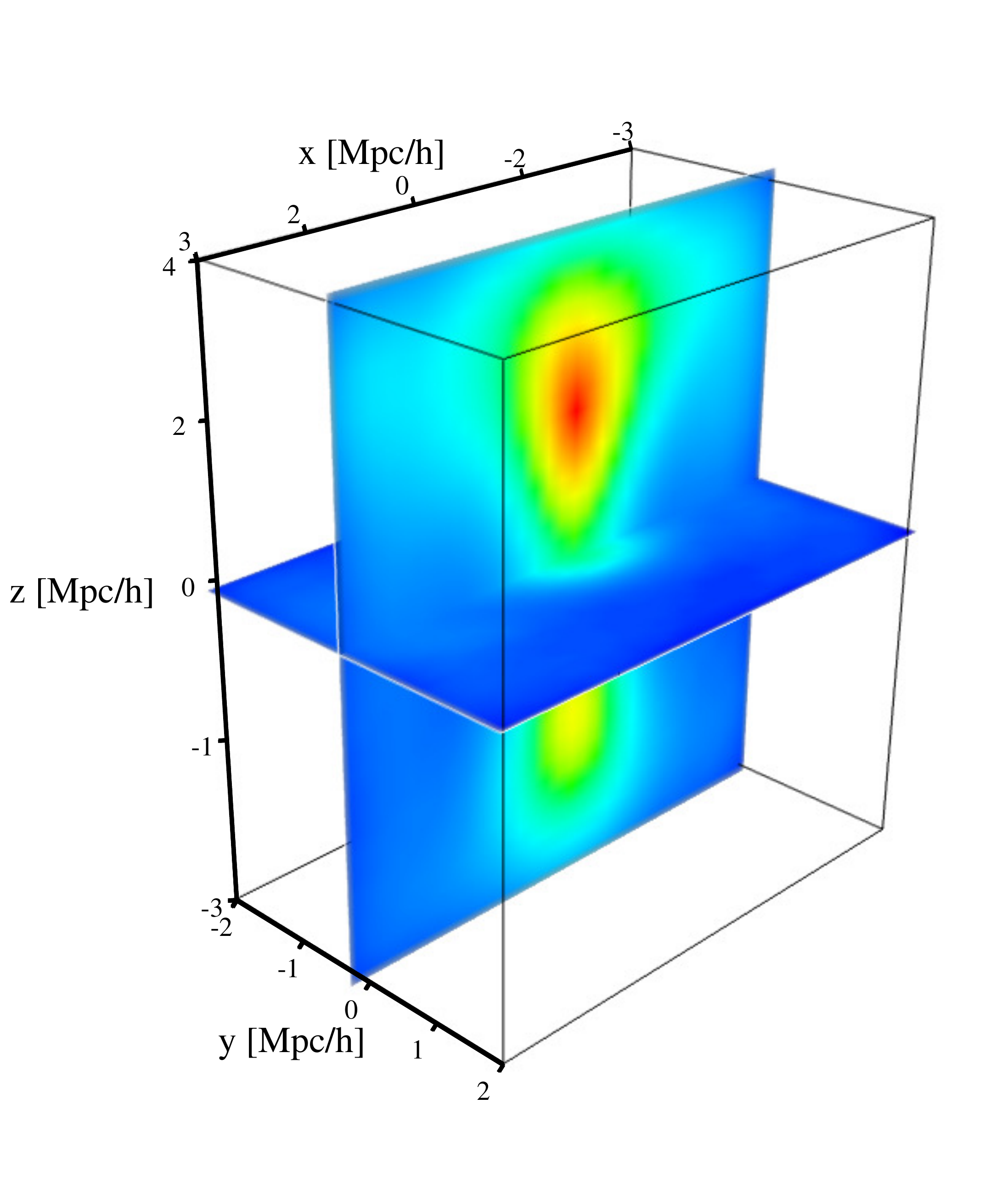}
\caption{3D structure of the neighbourhood of filaments at redshift zero. The galaxy number counts in the frame of the saddle for masses in the range $10^9$ to $10^{12} M_\odot$ ({\sl left}) are shown together with two 2D cross sections, longitudinal and transverse, of the filament at the saddle ({\sl right}).
The flattened flaring away from the saddle  reflects the co-planarity of filamentary bifurcation within the wall. The top-bottom asymmetry
reflects the orientation of the skeleton.}
\label{fig:3d_framesaddle}
\end{figure*}

\subsection{Galaxy properties}
\label{sec:galaxies}

The identification of galaxies is performed using the most massive sub-node method~\citep{Tweed2009} of the AdaptaHOP halo finder~\citep{Aubert2004} operating on the distribution of star particles with the same parameters as in~\cite{Dubois2014}.
Only structures with a minimum of $N_{\rm  min}=100$ star particles are considered, which typically selects objects with masses larger than  $2 \times 10^8 \, \rm M_\odot$. 
For each redshift output analysed in this paper ($0.05 <$ redshift $< 2$) catalogues containing up to $\sim 350 \, 000$ haloes and $\sim 180 \, 000$ galaxies are produced.

For each galaxy, its $V/\sigma$, stellar rotation over dispersion, is extracted from the 3D distribution of velocities.
This is meant to provide a kinematic proxy for morphology.
The total angular momentum (spin) of stars is first computed in order to define a set of cylindrical spatial coordinates ($r$, $\theta$, $z$), with the $z$-axis oriented along the spin of galaxy. The velocity of each individual star particle is decomposed into cylindrical components $v_{r}$, $v_{\theta}$, $v_z$, and the rotational velocity of a galaxy is $V=\bar v_{\theta}$, the mean of $v_{\theta}$ of individual stars. 
The average velocity dispersion of the galaxy $\sigma^2=(\sigma_{r}^2+\sigma_{\theta}^2+\sigma_z^2)/3$  is computed using the velocity dispersion of each velocity component $\sigma_{r}$, $\sigma_{\theta}$, $\sigma_z$.

\subsection{Saddle frame identification}

\label{sec:skeleton}
In order to quantify the position of galaxies relative to the cosmic web, a geometric three-dimensional ridge extractor called  {\sc \small  DISPERSE} \footnote{ The code {\sc \small  DISPERSE}, which stands for Discrete-Persistent-Structure-Extractor  algorithm is publicly available at the following URL \href{http://www.iap.fr/users/sousbie/disperse/}{http://www.iap.fr/users/sousbie/disperse/}.}~\citep{Sousbie2011,Sousbie2012} 
is run on the full volume gas density distribution over $512^3$ cells with a 3$\sigma$ persistence threshold. This density distribution is smoothed with a Gaussian kernel with smoothing length of $0.8$ comoving Mpc/$h$.
The orientation and distribution of galaxies can be measured relative to the direction of the closest filament's segment.  In particular, the code identifies saddle points along those filaments.
This is a costly method to identify saddle points, but it provides us with a local preferred  polarity in the frame of the density Hessian (positively towards
the larger of the two maxima).
It was checked that the distributions presented below are relatively insensitive to the choice of  smoothing length (see Appendix~\ref{sec:validation}).
It was also checked there that these results do not show a strong dependence on the tracer (dark matter or gas density) used to compute the skeleton.

\section{Saddle stacks in 2D and 3D}
\label{section:3Dgalaxies}

\begin{figure*}
\centering
\includegraphics[width=0.95\textwidth]{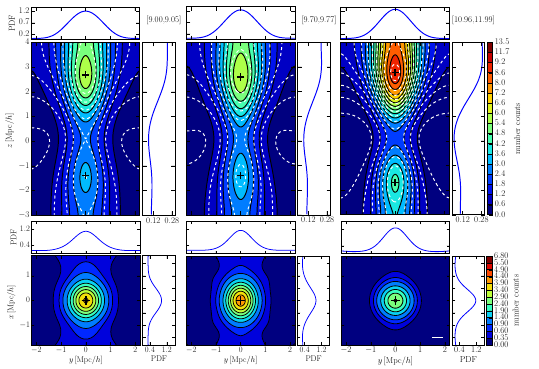}
\caption{The galaxy number counts at redshift zero in the frame of the saddle for low ({\sl left}), intermediate ({\sl middle}) and high ({\sl right}) stellar mass bins (see text for definition), as labeled (in square brackets), in the
longitudinal ({\sl top}) and transverse ({\sl bottom}) planes at the saddle. The vertical axis on top panels corresponds to the direction of the skeleton at the saddle (upwards toward the node with the highest density), while the horizontal axis corresponds to the major principal axis in the transverse direction.
The sub-panels on the top and on the right of each panel show the marginalised 1D distributions along respective axes.
The white dashed contours represent the galaxy number counts with the horizontal axis corresponding to the minor principal axis in the transverse direction at the saddle.
The black crosses represent the peaks in galactic density on axis and the white horizontal line represents the smoothing length used in the analysis.
The projection is carried over $ \pm 1$ Mpc/$h$ for the longitudinal slice and $ \pm 0.75$ Mpc/$h$ away from the saddles transversally.
The strength of the gradient of the galaxy number density changes with stellar mass. As expected, the high mass galaxies are more tightly clustered near the filament axis and near nodes ({\sl right panel}) compared to their low mass counterparts ({\sl left} and {\sl middle panels}).
}
\label{fig:framesaddle_counts_long_trans}
\end{figure*}

With the aim of studying the geometry of the galaxy distribution around filaments, stacking centred on the saddle points of filaments is applied.
When stacking, two different strategies are explored. First, stacks are produced centred on the saddle,
and physical properties of galaxies are binned as a function of transverse and longitudinal distances away from the skeleton.
These properties are also stacked in 3D in the local frame set by the direction of the filament at the saddle and the 2D inertia tensor in the plane perpendicular to the filament.
The former method avoids the flaring induced by the drift of the curved filaments away from the saddle,
only associate one saddle to each galaxy and stacks azimuthally,
while the latter one allows us to probe the transverse {anisotropic} geometry of filaments {at the saddle}.

\subsection{Azimuthally averaged stellar mass and  number density}
\label{section:2Dgalaxies}

Let us start by considering azimuthally averaged 2D maps in the frame defined by the saddle and its steepest ascent direction,
and study the cross sections of galactic number density and stellar mass in the vicinity of the saddle point. 
In order to infer the variation of galaxy properties beyond its stellar mass,  stellar mass will be fixed by considering 3 bins, defined as low (9.0 $\leq \log M_\star \leq$ 9.05), intermediate (9.69 $\leq \log M_\star \leq$ 9.75) and high  (10.93 $\leq \log M_\star \leq$ 11.99) stellar mass bins. These bins correspond to the first, middle and last 27-quantiles of the stellar mass distribution of all galaxies at a given redshift above the stellar mass limit of 10$^{9}$ \msun. Each of such constructed stellar mass bin contains $\sim$ 3500 galaxies.
The smoothing scale applied to the profiles is 0.4 Mpc/$h$\footnote{Changing the smoothing scale used to produce the maps to 0.2 and 0.8 Mpc/$h$  leads to qualitatively similar conclusions. The smoothing  impacts mostly the position of maxima in the transverse direction. At low values these tend to be offset from  the filament's axis because of the smoothing of the skeleton itself.}.


Figures~ \ref{fig:2D_curv_761_nb_allM_mass} and~\ref{fig:2D_curv_761_nb_allM_mass2} show the galactic number counts at low, intermediate and high stellar mass, and mean stellar mass for all galaxies above the stellar mass limit, respectively, at redshift zero in the frame of the saddle. In that frame, the vertical axis corresponds to the distance from the saddle point along the skeleton, upwards towards the densest node, while the horizontal axis corresponds to the transverse direction.
Note that the length of filaments is not constant, however its distribution is quite narrow with median length of $\sim$ 5.5 Mpc/$h$ at redshift zero (see Appendix~\ref{sec:filaments}, Figure~\ref{fig:fil_len}).
Iso-contours clearly display a dependence both on the radial distance from the saddle point and the orientation w.r.t. the filament's direction.
At fixed distance from the saddle point, the number of galaxies is enhanced in the direction of the filament, i.e. they are more clustered in the filaments than in the voids.
The  gradient of the number density of galaxies  is also found to change with stellar mass. The high mass galaxies are more tightly clustered near the filament axis and tend to be  further away from saddles along the filament compared to their low mass counterparts.
Saddle points are, as expected, local minima of both galaxy number counts and stellar mass in the direction along the filament towards the nodes, and local maxima in the perpendicular direction. Thus, galaxies in filaments tend to be more massive than galaxies in voids and within  filaments, while the stellar mass of galaxies increases with increasing distance from the saddle point in the direction toward nodes.
This effect is stronger in the direction perpendicular to the filament, where the relative variation of the mean stellar mass is about a factor of 2 higher compared to that along the filament.

The mass gradients shown on Figure~\ref{fig:2D_curv_761_nb_allM_mass2} can be qualitatively understood within peak and excursion set theories \citep[see Section~\ref{sec:theory} and][]{Codis2015a,Musso2018}.

\subsection{3D stacks of stellar mass and number density }
\label{subsection:3Dnb_mass_long}

Let us now investigate the 3D structure of the neighbourhood of filaments
by stacking galaxies relative to a  3D-oriented local reference frame, with its origin  defined by the position of the saddle point and its axes defined as follows: the $z$-axis corresponds to the direction of the filament at the saddle, and the $x$- and $y$-axes represent major and minor principal axes of the inertia tensor in the plane perpendicular to the filament axis at the saddle point, respectively\footnote{In practice, the 2D inertia tensor is computed by considering galaxies within $\pm$ 1 (Mpc/$h$) around the saddle point and projected into the plane perpendicular to the filament and passing through the saddle. Note that changing the volume of the considered region within a factor of  a few does not have a strong impact on   orientation. 
}. 

In order to increase the signal-to-noise ratio, galaxies are stacked by flipping them with respect to the filament axis to produce longitudinal cross sections, and with respect to both
principal axes of the inertia tensor in the case of transverse cross sections.

The 3D distribution of galaxies in such defined frame is shown in Figure~\ref{fig:3d_framesaddle} (left panel) together with planes representing 2D cross sections, longitudinal and transverse, as used in the analysis (right panel). In practice, individual cross sections are obtained by projecting galaxies within $\pm$ 1 Mpc/$h$ and $\pm$ 0.75 Mpc/$h$ from the plane passing through the saddle point for longitudinal and transverse cross sections, respectively.
Note the flaring near the nodes which arises because the typical saddle is flattened (the two negative eigenvalues of the Hessian differ, while the corresponding eigenvectors are aligned when stacking), 
and the Hessian remains correlated away from the saddle. Correspondingly, the skeleton bifurcates within that plane \citep{Pogosyan2009,connectix}. The top-bottom asymmetry reflects the fact that higher density contours 
are drawn near the more prominent peak (which is traced by the orientation of the skeleton).

As in the case of azimuthally averaged cross sections,  three stellar mass bins are defined as low (9.0 $\leq \log M_\star \leq$ 9.05), intermediate (9.7 $\leq \log M_\star \leq$ 9.77) and high (10.96 $\leq \log M_\star \leq$ 11.99) stellar mass bins, containing $\sim$ 10\,000 and $\sim$ 1000 galaxies, for longitudinal and transverse cross sections, respectively. The upward direction along $z$-axis corresponds to the direction of the node with highest density, and the smoothing scale applied to the profiles is 0.4 Mpc/$h$, as previously.

\begin{figure*}
\centering
\includegraphics[width=0.33\textwidth]{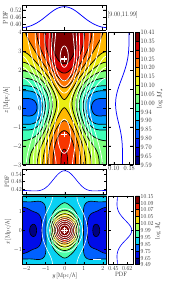}
\includegraphics[width=0.33\textwidth]{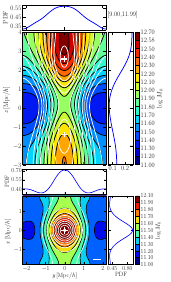}
\includegraphics[width=0.33\textwidth]{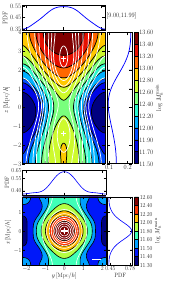}
\caption{Mean galaxy stellar mass ({\sl left}), sub-halo mass ({\sl middle}) and host halo mass ({\sl right}) in the frame of the saddle for masses in the range $10^9$ to $10^{12} M_\odot$ at redshift zero, in the longitudinal ({\sl top panels}) and transverse ({\sl bottom panels}) planes at the saddle. The vertical axis corresponds to the direction of the skeleton at the saddle (upwards toward the node with the highest density), while the horizontal axis corresponds to the major principal axis in the transverse direction.
The white contours correspond to the galaxy number counts with the horizontal axis corresponding to the major principal axis in the transverse direction at the saddle.
The white cross represent the peak in galactic density on axis. More massive galaxies are further away from the saddle than the low mass population in the longitudinal direction, while they are closer to the saddle transversally. As expected, more massive galaxies are also residing in more massive halos. Note in particular that the iso-contours of stellar mass are 
very similar to those of sub-halo mass, while the iso-contours of host halo mass, the shape of which differ from the two others, show much more 
resemblance to the iso-contours of density (as discussed in Section~\ref{section:discussion}).
The peak of maximum mass is further away from the saddle than the counts.
}
\label{fig:framesaddle_mass_long_trans}
\end{figure*}

 The cross sections of galactic number counts, stellar mass, specific star formation rate \ssfr$={\rm SFR}/M_{\rm *}$, where SFR is computed over a timescale of 50 Myr, \vsig and age will be studied in the vicinity of the saddle.
Figures~\ref{fig:framesaddle_counts_long_trans} and~\ref{fig:framesaddle_mass_long_trans} show the galaxy number counts in three different stellar mass bins, and mean stellar mass for all galaxies above the stellar mass limit, respectively, at redshift zero in the longitudinal (top panels) and transverse (bottom panels) planes in the frame of the saddle.
Once again, iso-contours clearly depend on both the radial distance from the saddle and the orientation w.r.t. the filament's direction.
Galaxies are found to be more clustered in  filaments than in  voids at all masses, i.e. at fixed distance from the saddle
point, the number of galaxies is enhanced in the direction of the filament. What changes with  stellar mass is the behaviour of the gradients
with the most massive galaxies being more tightly clustered near the filament
axis compared to their lower mass counterparts.
As in the case of azimuthally averaged cross sections, mass gradients seen in Figure~\ref{fig:framesaddle_mass_long_trans} (left panels) can be also understood in the context of constrained random field and excursion set theory, as discussed in Section~\ref{sec:theory}.

Interestingly, the distribution of most massive galaxies around the saddle points in the transverse direction is axisymmetric up to the distance of $\sim 1$ Mpc$/h$, while the iso-contours of lower mass galaxies are more flattened (in the direction of $x$-axis corresponding to the major axis of the inertia tensor in the transverse cross section) and extended to larger distances from the saddle.  This behaviour is a manifestation of the mass dependence of galaxy's connectivity: higher mass galaxies in denser environments are expected to be fed by numerous filaments while lower mass galaxies are typically embedded in a single filament \citep{connectix}.

\subsection{Longitudinal and transverse  \ssfr cross sections}
\label{subsection:3Dssfr_long}

\begin{figure*}
\centering
\includegraphics[width=0.83\textwidth]{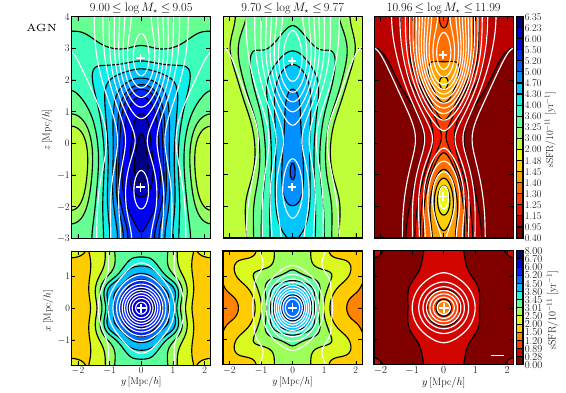}
\includegraphics[width=0.83\textwidth]{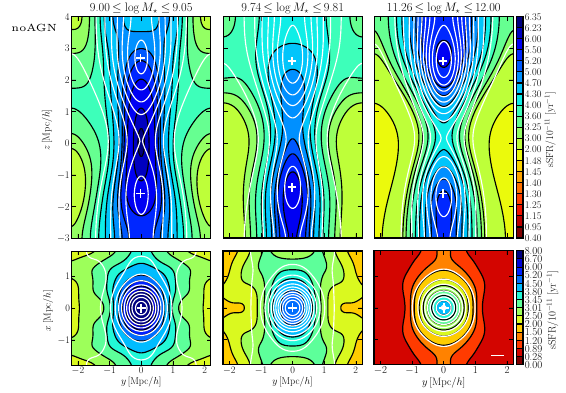}
\caption{Mass-weighted \ssfr in the frame of the saddle at redshift 0 for low ({\sl left}), intermediate ({\sl middle}) and high ({\sl right}) stellar mass bins, as labeled, in the
longitudinal and transverse planes at the saddle, in \hagn ({\sl topmost panels}) and \hnoagn ({\sl bottommost panels}). The vertical axis corresponds to the direction of the skeleton at the saddle (upwards toward the node with the highest density), while the horizontal axis corresponds to the major principal axis in the transverse direction at the saddle.
The white contours and the white crosses correspond to the galaxy number counts and the peak in galactic density on axis, respectively. The saddle represents maximum of \ssfr in transverse direction at all masses and regardless of the presence of the AGN feedback. What does change is the star formation activity in particular of the most massive galaxies, where AGN feedback substantially reduces the values of \ssfr. Moreover, note that at high mass end, the \ssfr iso-contours are modified by AGN feedback in the vicinity of the densest node, such that in the longitudinal direction away from the saddle, the maximum of \ssfr is off-set from the densest peak.
Overall, the \ssfr iso-contours display a stellar mass dependence in the longitudinal direction in that at low mass (resp. high mass) \ssfr is maximum (resp. minimum) at the saddle and it decreases (resp. increases) in the direction towards the nodes.
}
\label{fig:3D_ssfr_long}
\end{figure*}

Let us now focus on specific star formation rates.
Figure~\ref{fig:3D_ssfr_long} (top row) shows the mean stellar mass-weighted \ssfr at redshift zero in \hagn.
Iso-contours display qualitatively similar behaviour in all stellar mass bins in the direction perpendicular to the filament, for which the saddle point represents the maximum of \ssfr.
In the direction along the filament, the behaviour is more complex: at high stellar mass \ssfr increases with increasing distance from the saddle towards the nodes, but while the maximum of \ssfr overlaps with the position of the low density node, it is located closer to the saddle in the direction of the densest node, as will be discussed below. 
The \ssfr then decreases in this direction in the vicinity of the node and beyond. With decreasing stellar mass, the maximum of \ssfr  
moves closer to the saddle point, until it overlaps with the saddle point for lowest stellar mass bin.

A general trend of  decreasing \ssfr with increasing stellar mass is clearly recovered, with most massive galaxies having their \ssfr substantially reduced in particular in the vicinity of the densest node, where the average \ssfr value can be up to 10-times lower compared to their low mass counterparts.
Indeed  AGN feedback is   an important ingredient for the formation of the more massive galaxies, suppressing star formation so as to reproduce the observed high-end of the galaxy luminosity function.
By comparing the iso-contours of mean stellar mass-weighted \ssfr in \hagn and \hnoagn (bottom row of Figure~\ref{fig:3D_ssfr_long}), the two main
specific consequences of AGN feedback can be identified. First, and not surprisingly, when AGN feedback operates, the overall \ssfr is reduced, mostly in the high stellar mass bin (the mean \ssfr in the highest stellar mass bin changes by a factor of $\sim$ 3, while in the lowest stellar mass bin, it remains  $\sim$ 1.15).
Secondly, AGN feedback modifies the shape of \ssfr iso-contours. This effect is most prominent amongst most massive galaxies\footnote{Note that the highest stellar mass bin  is not identical in the two simulations. This is due to the difference in the stellar mass distributions, such that at high stellar mass end, there are more galaxies in \hnoagn than in \hagn that also tend to be more massive \citep[see also][]{Beckmann2017}. However, considering the same stellar mass bins does not impact our results.
Another difference is in the halo-to-stellar mass relation, especially at the high mass end. 
It was checked that the medians of halo masses in the highest stellar mass bin considered in this work  are comparable in both simulations.} in the vicinity of the densest node that represents the maximum of the \ssfr in the direction along the filament from the saddle when AGN feedback is absent. A similar effect is seen at low and intermediate stellar mass, albeit less pronounced.
Overall, the reduced star formation activity of galaxies due to AGN feedback in the densest environment translates into an offset of the maximum of the mean stellar mass-weighted \ssfr away from the node.
This clearly demonstrates the importance of the AGN feedback and its ability not only to reduce the star formation activity of individual objects, but also to modify their distribution on larger scales in the vicinity of high density regions such as nodes, corresponding to galaxy groups and clusters, consistently with our findings of AGN feedback being most efficient near nodes at high stellar mass (see Appendix~\ref{sec:AGN_noAGN_ssfr}).
%

\begin{figure*}
\centering
\includegraphics[width=\columnwidth]{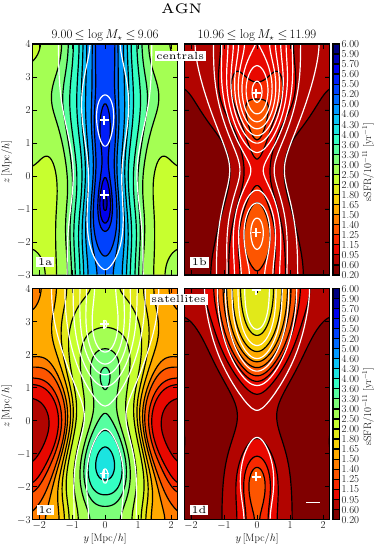}
\includegraphics[width=\columnwidth]{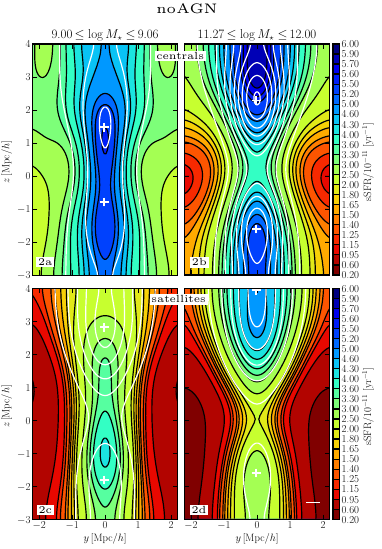}
\caption{Mass weighted \ssfr in the frame of the saddle at redshift 0 for low and high stellar mass bins, as labeled, in the longitudinal and transverse planes at the saddle, shown for centrals ({\sl top row}) and satellites ({\sl bottom row}) separately with ({\sl left}) and without ({\sl right}) AGN feedback.
The vertical axis corresponds to the direction of the skeleton at the saddle (upwards toward the node with the highest density), while the horizontal axis corresponds to the major principal axis in the transverse direction at the saddle.
The white contours and the white crosses correspond to the galaxy number counts and the peak in galactic density on axis, respectively.
AGN feedback has the strongest impact on high mass centrals and in the vicinity of the densest node (compare panel 1b with 2b and with panel 1d), where it modifies the shape of the \ssfr iso-contours as already noticed for the entire high mass population (see Figure~\ref{fig:3D_ssfr_long}).
At low stellar mass, satellites are generally less star-forming compared to centrals, but note also that the \ssfr iso-contours of centrals and satellites are also different. For satellites, the maximum of \ssfr is located between the saddle and the peak in the direction along the filament towards the densest node (compare panel 1a with panel 1c or panel 2a with panel 2c). 
}
\label{fig:3D_curv_ssfr_cen_sat}
\end{figure*}

\subsection{Centrals and satellite differential counts}
\label{subsection:satellite}

\begin{figure*}
\centering
\includegraphics[width=0.75\textwidth]{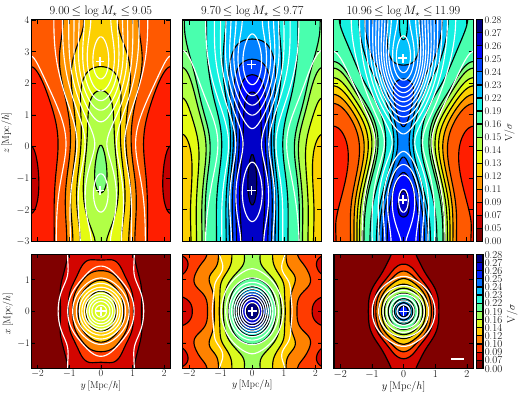}
\includegraphics[width=0.75\textwidth]{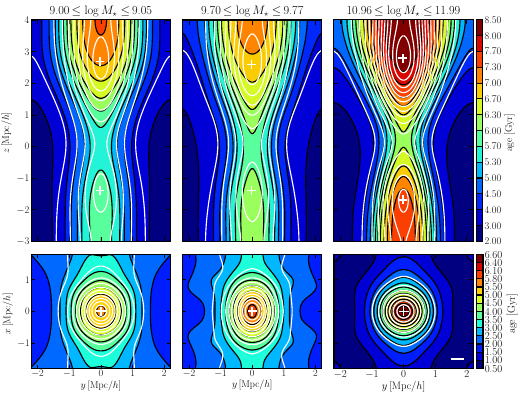}
\caption{Stellar mass weighted \vsig ({\sl topmost panels}) and age ({\sl bottommost panels}) for low ({\sl left}), intermediate ({\sl middle}) and high ({\sl right}) stellar mass bins as labeled, in the longitudinal and transverse planes at the saddle. The vertical axis corresponds to the direction of the skeleton at the saddle (upwards toward the node with the highest density), while the horizontal axis corresponds to the major principal axis in the transverse direction at the saddle.
The white contours and the white crosses correspond to the galaxy number counts and the peak in galactic density on axis, respectively.
The behaviour of the number density 
of galaxies changes with stellar mass for both physical properties, but much more dramatically for \vsig. The shape of iso-contours are qualitatively different, while the maximum of \vsig along the filament in the upward direction is located in between the saddle point and the density peak, age increases with increasing distance away from the saddle towards the nodes and beyond. Transverse gradients are similar, both in terms of shape of iso-contours and in that the saddle point is maximum for both quantities in radial direction.
}
\label{fig:framesaddle_vsigage_long}
\end{figure*}

In order to gain a better understanding of what processes regulate \ssfr of galaxies in their anisotropic environment,  galaxies are next split into centrals and satellites (respectively the most massive galaxy within 10 per cent the virial radius of halo, or a subhalo). Making this separation is further motivated by more straightforward comparison with theoretical prediction of \cite{Musso2018} (that is strictly applicable to central galaxies alone, as the effect of the large-scale tidal field on the low mass objects is not accounted for).
Figure~\ref{fig:3D_curv_ssfr_cen_sat} shows stellar mass-weighted \ssfr for centrals (top panels) and satellites (bottom panels) separately in both simulations, \hagn (leftmost panels) and \hnoagn (rightmost panels) in low and high stellar mass bins.
Not surprisingly, the low mass end is dominated by the population of satellites, while central galaxies dominate the highest stellar mass bins.
What is more interesting is the distinct response of centrals and satellites in terms of their \ssfr as a function of the exact position within the cosmic web (in both \hagn and \hnoagn) and more surprisingly,
the distinct impact of AGN feedback on the \ssfr of these two populations.

AGN feedback seems to have a stronger impact on centrals which are closer to the denser node (compare panel 1b with 2b and 1d with 2d).
At high mass, AGN feedback quenches much less efficiently star formation in satellites than it does in centrals (compare panel 1b with 1d), where it distorts the shape of the \ssfr iso-contours in the vicinity of the denser node. 
High mass satellites seem to feel the impact of both the AGN feedback and environmental processes, in particular in dense regions, but less so than the centrals (compare panel 1d with 2d). 
A possible explanation for massive satellites being less affected by the AGN feedback (compared to centrals at the same stellar mass) could be the tidal influence of their main halo \citep{Hahn2009} which  reduces  accretion and merger rate onto the satellite. As mergers  trigger bursts of AGN activity, this induces less star formation.

At low stellar mass, as expected, AGN feedback does not seem to have a strong impact on the sSFR of both satellites and centrals (compare panel 1a with 2a and 1c with 2c). 
At low stellar mass, \ssfr iso-contours are different for satellites and centrals: i) satellites have lower sSFR compared to centrals of the same mass, in the direction both perpendicular to the filament, and along the filament towards the nodes, and ii) the shape of sSFR iso-contours is different for satellites and centrals, in particular in the vicinity of denser node, in that for satellites, the sSFR reaches its maximum before reaching the densest node in the direction along the filament (compare panel 1a with 1c or 2a with 2c). Presumably, satellite specific processes, such as e.g. strangulation, are driving this difference\footnote{Strangulation \citep{larsonetal1980}, together with mergers \citep{Toomre1972}, are traditionally considered as  group-specific processes impacting star formation activity of satellites. Other environmental quenching processes, mostly operating in clusters include galaxy harassment \citep{mooreetal1996} or ram pressure stripping of gas \citep{Gunn1972}. However, in this work, we are not attempting to address the processes impacting satellite population in particular.}.
Note that the \ssfr contours for massive centrals in the \hnoagn simulation (see panel 2b) are, as expected, in qualitative agreement with the dark matter accretion predicted by~\cite{Musso2018} (see Section~\ref{section:discussion} for a more detailed discussion).

\subsection{Longitudinal and transverse kinematic/age  cross sections}
\label{subsection:3Dlinematics_long}

Let us finally focus on the kinematics, quantified by the ratio of rotation to dispersion - dominated velocity, \vsig, and the age of galaxies in the frame of the saddle. The observational proxies of these quantities would be morphology and colour, respectively. Higher \vsig typically characterises disc dominated morphologies, while lower \vsig indicates the presence of a substantial bulge component. The age of galaxies corresponds to the mean ages that are given by the mass-weighted age of star particles belonging to each galaxy.
Figure \ref{fig:framesaddle_vsigage_long} shows iso-contours of \vsig (top panels) and age (bottom panels) as a function of stellar mass at redshift zero. Again, the contours exhibit both radial and angular gradients w.r.t. the saddle point.
At all stellar mass bins, galaxies tend to have higher \vsig in the vicinity of the saddle point that decreases in the orthogonal direction away from the saddle, while in
the direction along the filament towards the nodes it first increases, reaches its maximum before getting to the densest node and decreases afterwards. This effect is strongest for highest mass galaxies.
In terms of quantitative comparison of \vsig at different stellar mass, galaxies in the lowest stellar mass bin have the lowest \vsig, while intermediate mass galaxies show the largest \vsig values. \vsig of the most massive galaxies is lower compared to intermediate stellar masses, but higher than at lowest stellar mass end.   
This can be explained by the presence of few massive disc dominated galaxies present in the \hagn simulation and higher fraction of ellipticals at low mass end compared to observations. 
Indeed, as shown in \cite{Dubois2016}, the maximum probability of finding discs in \hagn is in the stellar mass range of $10^{10} -10^{11} M_\odot$. 

Similarly, age gradients display clear radial and angular dependence w.r.t. the saddle point at all stellar mass bins, however, with qualitatively different behaviour. In the transverse direction, saddle point is still maximum of the age at all stellar mass, while in the direction along the filament away from the saddle, age increases all the way beyond the node. Interestingly, in this aspect, age gradients are similar to stellar mass gradients with the oldest and most massive galaxies being
located closer to the node in the direction of the filament, and in the vicinity of the filament in the orthogonal direction.
This is consistent with the redshift evolution of the stacks as discussed now.

\begin{figure*}
\centering
\includegraphics[width=0.9\textwidth]{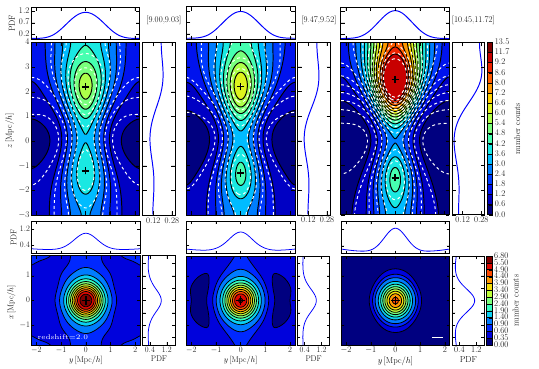}
\includegraphics[width=0.9\textwidth]{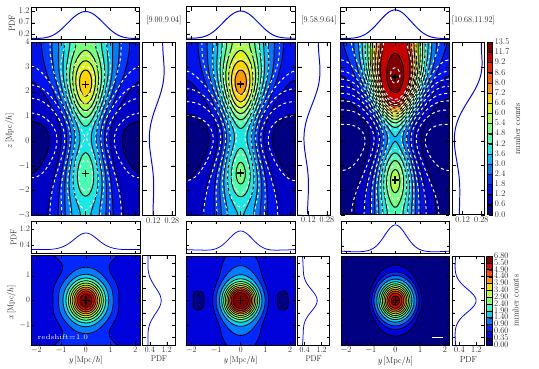}
\caption{Redshift evolution of the galaxy number counts in the frame of the saddle, in the longitudinal and transverse planes at the saddle. Low ({\sl left column}), intermediate ({\sl middle column}) and high ({\sl right column}) stellar mass bins are shown at redshifts 2 ({\sl top-most panels}) and 1 ({\sl bottom-most panels}), respectively. The white dashed contours represent the galaxy number counts with the horizontal axis corresponding to the minor principal axis in the transverse direction at the saddle. The corresponding redshift zero maps are shown on Figure~\ref{fig:framesaddle_counts_long_trans}. High mass galaxies are more clustered near the filaments and nodes at all redshifts considered compared to their lower mass counterparts. Note that as galaxies grow in mass with time, they follow the global flow of matter, reflected by the increased distance between the saddle point and two respective nodes at lower redshift. 
}
\label{fig:3D_1D_curv_redshift_nb}
\end{figure*}

\begin{figure*}
\centering
\includegraphics[width=0.7\textwidth]{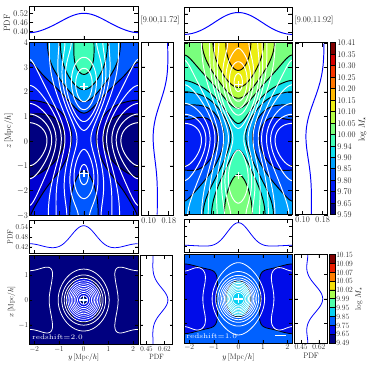}
\caption{Mean galaxy stellar mass in the frame of the saddle, in the longitudinal and transverse planes at the saddle, for all masses in the range $10^{9.0}$ to $10^{12.0} M_\odot$ at redshifts two ({\sl left column}) and one ({\sl right column}). The redshift zero maps are shown on Figure~\ref{fig:framesaddle_mass_long_trans}. 
At a given mass, the corresponding (coloured) contours get further away from the filament axis with cosmic time.
Transverse cross sections ({\sl bottom panels}) of number counts (white contours) become more elongated with decreasing redshift, while longitudinally ({\sl top panels}), they are further away from the saddle at lower redshift: the filaments become more elliptical and thicken with cosmic time (see Figure~\ref{fig:filament_thickness} for quantitative estimate of this effect). 
}
\label{fig:3D_curv_redshift_allM_mass}
\end{figure*}

\section{Redshift evolution}
\label{section:redshift}

Let us now examine the evolution of galaxy properties with redshift.
When comparing different epochs one may either consider the fate of a given set of galaxies, or quantify the cosmic evolution of the galactic population as a whole.

Figure~\ref{fig:3D_1D_curv_redshift_nb} shows galaxy number counts in low (left column), intermediate (middle column) and high (right column) stellar mass bins at redshifts two (topmost rows) and one (bottommost rows)\footnote{The skeleton and stellar mass bins are constructed as for redshift zero, see Sections~\ref{sec:skeleton} and~\ref{subsection:3Dnb_mass_long}, respectively. Consequently, the stellar mass bins are not identical at different redshifts, but they still contain comparable number of galaxies.}, while Figure~\ref{fig:3D_curv_redshift_allM_mass} shows the mean stellar mass of the entire population above the mass limit at these redshifts, as indicated\footnote{Note that these cross-sections are in qualitative agreement with azimuthally averaged counterparts (see Appendix~\ref{sec:azimuthal}).}. The corresponding redshift zero maps are shown on Figures~\ref{fig:framesaddle_counts_long_trans} and \ref{fig:framesaddle_mass_long_trans}, respectively.

At each redshift, more massive galaxies are more tightly clustered in the filaments than in the voids, and near the nodes than near the saddles.
Part of this redshift evolution is simply due to the mass evolution of objects. In other words, one could fix the level of non-linearity by considering mass bins that evolve with redshift following the non-linear mass for instance and then consider the residual redshift evolution. This procedure would allow to focus on the same class of objects across redshifts.

On Figure~\ref{fig:3D_1D_curv_redshift_nb}, one can follow the progenitors of a given class of objects by fixing the level of non-linearity which is equivalent to move approximatively along the diagonal (by adding Figure~\ref{fig:framesaddle_counts_long_trans}), i.e. to focus on less massive objects at high redshift.
As galaxies grow in mass, i.e. as non-linear gravitational clustering proceeds (the local dynamical clocks being set by inverse square root of the local density), they also become more concentrated towards the filaments and nodes (see Appendix~\ref{sec:cubic}).
For instance, comparing the bottom right transverse cross-section at redshift one and zero (from Figure~\ref{fig:framesaddle_counts_long_trans}), the vicinity of the saddle is
less populated by massive objects as these have drifted towards the nodes.
This redshift evolution is consistent with the global flow of galaxies first towards the filaments and then along them  (as  quantified kinematically in Appendix~\ref{sec:vel}), 
and with the fact that galaxies accumulate mass with cosmic time.

\begin{figure}
\centering
\includegraphics[width=0.9\columnwidth]{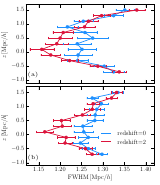}
\caption{Thickness of the filaments, defined as the FWHM of the Gaussian fit of the transverse galaxy number counts profiles marginalised over $x$- ({\sl panel a}) and $y$-axis ({\sl panel b}) at different positions along the filament's direction ($z$-axis on the longitudinal cross sections) at redshifts 2 and 0, in red and blue, respectively, for all galaxies in the mass range $10^9$ to $10^{12} M_\odot$. The transverse projections are carried over 0.2 Mpc/$h$ longitudinally (along the $z$-axis). As previously, the upward direction along the $z$-axis corresponds to the direction of the skeleton at the saddle toward the node with the highest density. When considering the entire population of galaxies, the cross sections of filaments in the vicinity of the saddle point (at $z=0$ Mpc/$h$) grow with time. For the sake of clarity,  only  measurements at redshifts zero and two are shown, however, their redshift evolution is consistent throughout. Note also that the widths  are computed in comoving coordinates: the growth at low redshifts is much stronger in physical coordinates. See also Appendix~\ref{sec:filaments} (Figure~\ref{fig:fil_thick}) for the thickness of the filaments and its redshift evolution at distances extending more faraway from saddle.
}
\label{fig:filament_thickness}
\end{figure}

For a population as a whole, in the close vicinity of the saddle, the breadth of the filament broadens with cosmic time as shown in Figure~\ref{fig:filament_thickness}, comparing the filament's thickness for all galaxies above the stellar mass limit at redshifts two and zero.
Specifically, the full width at half maximum (FWHM) of the transverse galaxy number counts profiles was computed at different positions along the filament's direction.
As argued in the next section, the measured increase of the filament's width with cosmic time is consistent with the theoretical expectations.

\begin{figure*}
\centering
\includegraphics[width=0.7\textwidth]{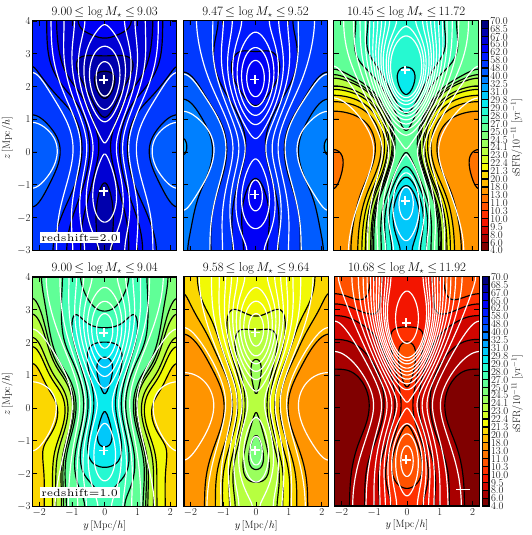}\\
\caption{Mass-weighted \ssfr in the frame of the saddle, in the longitudinal plane. Low ({\sl left column}), intermediate ({\sl middle column}) and high ({\sl right column}) stellar mass bins are shown at redshifts two ({\sl top}), and one ({\sl bottom}), respectively.
The redshift zero map is shown on the top panel of Figure~\ref{fig:3D_ssfr_long}.
\ssfr decreases with cosmic time at all stellar masses independently of the relative position w.r.t. the saddle. Interestingly, the peak of \ssfr drifts away from the densest node as a function of cosmic time or increasing mass (i.e. level of non-linearity).
}
\label{fig:3D_curv_redshift_ssfr}
\end{figure*}

Finally, Figure~\ref{fig:3D_curv_redshift_ssfr} shows the redshift evolution of stellar mass-weighted \ssfr.
Again, it is interesting to note that the global \ssfr traces the level of non-linearity of the collapse of structures:
at high redshift, low mass population (top left panel) has the highest sSFR, whereas the high mass low redshift population (bottom right panel) is the most quenched. This is also reflected in the position of maximum of \ssfr, which drifts with cosmic time i.e. with the level of non-linearity of the field.
The peak of \ssfr seems to occur further from the denser nodes towards the saddle as a function of cosmic time.
Hence, for the \ssfr at least two processes compete: advection with the main flow and
star formation activity which is impacted by the proximity to AGNs and the local dynamical timescale (but see Section~\ref{sec:spin} below).

\section{Theoretical predictions}\label{sec:theory}

Let us briefly present the theoretical framework which will allow us to interpret the measurements presented in Sections~\ref{section:3Dgalaxies} and~\ref{section:redshift}. This will involve predictions for 
dark matter  and halo density cross sections in the frame of the saddle, and  their  expected non-linear evolution with cosmic time.

\subsection{Constrained random fields}
For Gaussian cosmological initial conditions, peak theory \citep{BBKS} can be adapted  to predict the mean (total) matter density maps around saddles.
 Appendix~\ref{section:theory}  derives this mean initial matter distribution marginalised  to the constraint of  a saddle point of arbitrary geometry (height and curvatures) when the direction of the largest (positive) eigenvalue {of  the Hessian}, i.e the direction of the filament, is fixed together with its orientation. This last requirement is achieved by imposing that the coordinate of the gradient of the gravitational potential along the filament is always negative.
The resulting oriented   map of the density distribution around saddles is shown in Figure~\ref{fig:halo-mass-comparison} (left and middle panels). As expected, more mass is found close to the filament axis and in the direction of the most attractive potential well (towards the top of the map).
Figure~\ref{fig:halo-mass-comparison} (right panel) also presents the expected mass distribution of dark matter haloes within the frame of the saddle when the Press-Schechter threshold for collapse is decreased by the mean density
\citep[following the prescription described in][]{Codis2015a}.

\subsection{Expected redshift evolution} \label{sec:redshift}
Different approaches can be used to incorporate the non-linear evolution in the theoretical predictions, e.g. by doing a Zel'dovich boost of the mean density map predicted from excursion set theory, or by incorporating  the gravity induced non-gaussianity of the distribution using a perturbative approach as sketched in Appendix~\ref{section:theory}.
Both predict  that gravitational clustering
 distorts and enhances the contours of the matter density field within the frame of the saddle, with a scaling
proportional to  $\sigma(M_\star,$redshift), the mass- and redshift-dependent scale of non-linearity.
The net effect will depend on what is held fixed while stacking.
At fixed rareness, which is essentially  achieved when focusing on the more massive objects, filaments will collapse with cosmic time and therefore get thiner and more concentrated (see Figure~\ref{fig:zeldoboost}).
On the other hand, when the entire population of galaxies is considered at each redshift, filaments typically get thicker, because  less rare and therefore less connected and less biased objects form at low redshift and dominate the population.

While the realm of these predictions is limited (in redshift and range of tracers),
it nonetheless allows us to understand the trend at the level of gravity-driven processes,
and highlight by contrast the contribution of AGN or stellar feedback. 
We refer to  \cite{Codis2015a} (their Section~4) and \cite{laigle2015} (their Section~5) for predictions for the expected angular momentum and vorticity distributions and their evolution in the frame of the saddle, which will prove useful when discussing \vsig maps (and less directly \ssfr maps, which are sensitive to
the recent accretion of cold gas).

\section{Interpretation and discussion}
\label{section:discussion}

Let us now discuss the findings of Sections~\ref{section:3Dgalaxies} and~\ref{section:redshift} in the context of existing surveys and structure formation models (Section~\ref{sec:theory}).

\subsection{Complementary top-down approach to galaxy formation}
\label{subsec:topdown}

Let us start by putting the adopted approach and the results of this work in the classical context of  structure formation models. Traditionally, galaxy formation and evolution is studied in the hierarchical framework where galaxies are considered as evolving in (sub)-halos  possibly embedded in larger halos \citep[e.g.][]{Kauffmann1993,white1996}. 
Dynamically, this means that we can associate two typical timescales (or `clocks') to each encapsulated environment. 
 This approach is  justified in the well-established bottom-up scenario of structure formation. 
One can address the impact of the isotropic environment on the scales of halos, or equivalently the local density   (i.e. the trace of the Hessian of the gravitational potential)  while considering the merger tree history of individual halos (and thus galaxies residing within)\footnote{The local density is indeed strongly correlated with the group halo mass, 
as can be seen by comparing e.g. Figures~\ref{fig:framesaddle_mass_long_trans} and \ref{fig:3D_curv_761_dens}.}.
 Such scenario has proven quite successful in explaining many observed properties of galaxies, via the so-called  halo model
\citep{Sheth2002} -- in particular against isotropic  statistics (e.g. two point functions).   In this classical view the impact of the larger anisotropic scales set by the cosmic web  is ignored because it is assumed that these scales do not couple back down to galactic scales.  
Yet this view fails to capture e.g. spin alignments which are specifically driven by scale-coupling
to the cosmic web \citep{codis2015}, nor does it fully take into account how the  light-cone of a given galaxy 
is gravitationally sensitive to the larger scale anisotropies.

By contrast,  \cite{Musso2018}  recently investigated the impact of the large-scale anisotropic cosmic web on the assembly history of dark matter haloes within the framework of extended excursion set theory,
accounting for the effect of its large-scale tides. They derived the typical halo mass, typical accretion rate and formation time of dark matter haloes as a function of the geometry of the saddle.
These quantities were predicted to vary with the orientation and distance from saddle points, such that haloes in filaments are less massive than haloes in nodes, so that at equal mass they have earlier formation times and smaller accretion rates at redshift zero, the effect being stronger in the direction perpendicular to the filament.
These findings suggest that on top of the mass and local mean density, the tides of the larger scale environment   also impact haloes' properties through a third timescale.

The approach adopted here follows up and assesses specifically the impact of this  large-scale environment on {\sl galaxy} properties, and in particular the  top-down relevance of the imposed tides  (captured by the traceless part of the Hessian of the gravitational potential) on galaxy assembly. 
In other words the aim here is to identify properties of galaxies which are specific to their relative position within the saddle frame.
To do so, the analysis  is carried out at fixed stellar mass and quantified at additional fixed (sub)-halo mass and anisotropic density  (through the analysis of stacked re-oriented residual maps, see below), instead of the  conventional galaxy-halo-group mass isotropic perspective. This framework  does not invalidates  past results expressed in terms of group and halo masses -- which remain the dominant effect impacting galaxy formation, but
complements them at first or second order corrections\footnote{In fact one could indeed alternatively extend the classical framework by adding the larger-scale group distribution, i.e. the cosmic web  traced by dark matter halos  as a extra `hidden variable' driving galactic assembly. Below that scale, the statistics is isotropic, while beyond it one has to define how ensemble average should be carried.
The frame of its saddles is chosen here as a proxy for this web so as to be able to stack galactic distributions while taking its 
effect into account.}. Qualitatively the aim is to understand the impact of the stretching and twisting imposed by those tides above and below the impact of the density.  
As shown in Section~\ref{sec:redshift} it also provides as a bonus  a good understanding of the bulk  flows within that frame, which enlightens the geometry of filaments' iso-contours traced by galaxies at fixed mass or fixed cosmic age.

\subsection{Observational signature for the impact of the cosmic web}

The idea that galaxy properties, such as their stellar mass, colour or \ssfr  are also driven specifically by the anisotropy of the cosmic web has only recently started to be explored in observations \citep[e.g.][]{Eardley2015,Alpaslan2016,Tojeiro2017}.
Stellar mass and colour or \ssfr gradients have been reported at low \citep[e.g.][]{Chen2017}, intermediate \citep[$z \lesssim 0.25$;][]{Kraljic2018} and higher redshifts \citep[$z \sim 0.7-0.9$;][]{Chen2017,Malavasi2017,Laigle2018},
with more massive and/or less star forming galaxies being found closer to the filaments compared to their lower mass and/or higher star forming counterparts.
The focus in the present paper is on 2D and 3D cross sections at fixed stellar mass, allowing to explore more complex geometric environment of the filamentary network.  The (marginalised) 1D distributions (over distance along the filament) are in qualitative agreement with the above mentioned observed stellar mass and colour or \ssfr gradients w.r.t. filaments.
Marginalising over the distance perpendicular to the axis of the filaments yields gradients along the filament, such that at fixed orthogonal distance from the filament, more massive and/or less star forming galaxies are preferentially located in the vicinity of the node. Such a signature was found by \cite{Kraljic2018} in terms of red fractions, who reported the increasing fraction of passive galaxies with decreasing distances both to the filaments and nodes, with the dominant effect being the distance to the nodes.
These gradients should now be measured in the 3D distribution of galaxies inferred from large galaxy redshift surveys, such as e.g. SDSS \citep{York2000} or GAMA \citep{Driver2009,Driver2011}, providing a large statistical sample of galaxies and for which additional information about the properties of group halos is available.

In terms of redshift evolution of \ssfr,  note that while at redshift above one the \ssfr of galaxies increases in the direction along the filament away from the saddle and reaches its maximum near the node -- in the region where the density is typically highest, this maximum is {\sl shifted} away from the nodes towards the saddle at redshift one and below (top panels of Figure~\ref{fig:3D_ssfr_long} and Figure~\ref{fig:3D_curv_redshift_ssfr}). 
Qualitatively similar behaviour, known as the reversal of the star formation-density relation at high redshift, was tentatively identified in observations \citep[e.g.][but see e.g. \citealt{Patel2009,Popesso2014}, for contradictory results]{elbazetal2007,Cooper2008,Hwang2010}\footnote{\cite{elbazetal2007} specifically found  evidence of this reversal for massive galaxies, such that the \ssfr increases with increasing galaxy density at redshift $\sim$ one.}.
Overall, our results suggest that in order to understand the complex behaviour of galaxies' properties, one may need to take into account the large-scale environment where tides are expected to play an important role, beyond that of density.

Note finally that a possible reason for the recent  non-detection of \cite{Alam2018} and \cite{Paranjape2018b} with the SDSS resolution
 is that the ensemble average of the non-linearly evolved galactic properties predicted from angular averaged fields does not differ by much from 
the ensemble and angular  average of the non-linearly evolved galactic properties  from anisotropic  fields.
To a good approximation, angular-averaging and dynamical non-linear evolution commute,
 which has of course been the basis of the success of the spherical collapse model\footnote{This is in fact seen even at the level of the one-point function: one needs to invoke  a moving barrier \citep{2002MNRAS.329...61S}, i.e.  corrections to spherical collapse to match the measured mass function of dark halos.}.
One has to compute expectation in the frame of the filament to underline the differences, which is precisely the purpose of this paper.
 
\subsection{Inferred age, mass and counts statistics}
\label{subsec:musso}

The findings presented in this work,
based on the analysis of galaxy-related gradients in the frame of saddle, are in qualitative agreement with the  predictions of  \cite{Musso2018}  and those of Section~\ref{sec:theory}: the iso-contours of studied galaxy properties show dependence on both the distance and orientation w.r.t. the saddle point of the cosmic web.
Specifically, galaxies tend to be more massive closer to the filaments compared to voids, and inside filaments near nodes compared to saddles (Figures~\ref{fig:2D_curv_761_nb_allM_mass} - \ref{fig:framesaddle_mass_long_trans}).
Similarly and equivalently (given the duality between mass and cosmic evolution discussed in Appendix~\ref{sec:cubic}),  Figures~\ref{fig:3D_1D_curv_redshift_nb} - \ref{fig:filament_thickness} show that as galaxies grow in mass, they become more clustered near filaments and nodes with cosmic time, the width of the filaments narrows for a given mass bin, while the evolution of the entire population is consistent with broadening of the filaments, as expected from the theory of rare events \citep{Bernardeau1994}.
The number counts maxima are closer to the saddles than the stellar mass maxima as the former is dominated by the less-massive  and more-common population,  forming more evenly within the frame of the cosmic web, so that they have not had time to drift to the nodes. Consistently, older galaxies (Figure~\ref{fig:framesaddle_vsigage_long}) are preferentially located near the nodes of the comic web when comparing their distribution in the direction along the filaments, 
and in the vicinity of filaments in the perpendicular direction.
These age gradients are seemingly at odds with the formation time of haloes predicted by \cite{Musso2018}, where haloes that form at the saddle point assemble most of their mass the earliest. However, note that the formation time of haloes does not necessarily trace  galactic age as inferred from the mean age of the stellar population. Indeed, our findings reflect the so-called downsizing \citep{Cowie1996}  of both galaxies and haloes \citep[e.g.][]{Neistein2006,Tojeiro2017}, such that oldest galaxies tend to be most massive, and galaxies in high mass haloes are older (they formed their stars earlier).

Note finally that the theoretical predictions in \cite{Musso2018} are made at fixed halo mass, while the analysis presented so far in this work is performed at fixed stellar mass. However, the halo mass used in their study is  physically closer to a sub-halo mass than a host halo mass\footnote{The formalism adopted in \cite{Musso2018} does not   capture the strongly non-linear processes operating on satellite galaxies.}, and is therefore more strongly correlated with the stellar mass of galaxies which justifies further the qualitative comparison at this stage.   
As anticipated in Section~\ref{subsec:topdown}, additional fixed sub-halo mass and density will be taken into account through the analysis of residuals (see Section~\ref{sec:hidden_variable}).

\begin{figure*}
\centering
\includegraphics[width=0.9\textwidth]{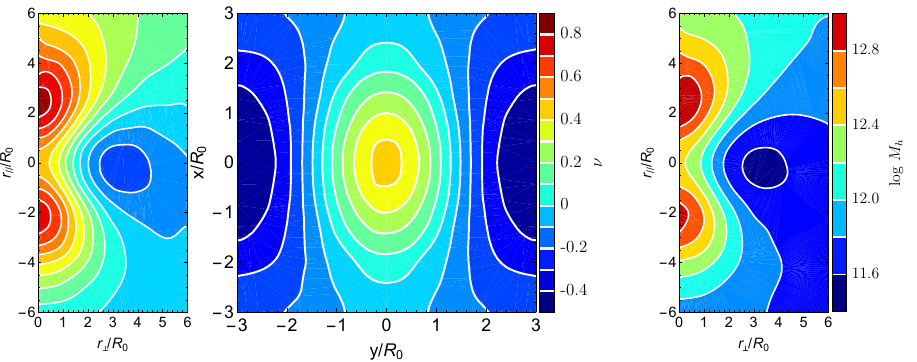}
\caption{Mean predicted maps of the dark matter distribution around a saddle point of arbitrary shape and height. The direction of the filament is fixed to be along the vertical axis for left and right panels (perpendicular to the plane of the figure in the middle panel) and   the top-bottom symmetry is broken by imposing that the most attractive peak is at the top. {\sl Left panel}: predicted distribution of the density fluctuation for Gaussian random fields  $\nu$ (in units of the variance). {\sl Middle panel}: corresponding transverse cross section with the same colour coding. {\sl Right panel}: corresponding log of the non-linear mass when the threshold for collapse in the Press-Schechter mass is decreased by the mean density obtained on the left panel. Its numerical counterpart measured in \hagn is shown on Figure~\ref{fig:mass-evolution} at low and high redshifts (see also  Figure~\ref{fig:zeldoboost} for a prediction). 
}
\label{fig:halo-mass-comparison}\label{fig:xy-plane-GRF}

\end{figure*}

\subsection{The impact of AGN feedback}
Relating the predicted specific accretion gradients of dark matter haloes to galaxies' observables requires some assumptions.
One can in principle translate dark matter accretion gradients into \ssfr gradients by considering the role of baryons in the accretion and feedback cycle.
In the current framework of galaxy formation and evolution, galaxies acquire their gas by accretion from the large-scale cosmic web structure. The average growth rate of the baryonic component can be related to the cosmological growth rate of dark matter haloes, from which follows that higher star formation rate corresponds to higher dark matter accretion rate, providing that the SFR follows the gas supply rate.
At high redshift, the vast majority of galaxies are believed to grow by acquiring gas from steady, narrow and cold streams \citep[e.g.][]{Keres2005,Ocvirk2008,Dekel2009}.
Using these arguments, it should follow that at high redshift, the stronger the accretion, the higher the \ssfr of galaxy. 
Such a scenario is consistent with the gradients of the dark matter accretion rates found by \cite{Musso2018}, where high mass haloes that form in the direction of the filament tend to have higher accretion rates than haloes with the same mass that form in the orthogonal direction.
This  qualitatively agrees with  the \ssfr gradients in the frame of saddle at high redshift (Figure~\ref{fig:3D_curv_redshift_ssfr}) and in the simulation {\sl without} AGN feedback (Figure~\ref{fig:3D_ssfr_long}) at redshift zero, where galaxies with highest \ssfr at fixed stellar mass tend to be located in the vicinity of the node in the direction along the filament,  and near the saddle in the orthogonal direction.

In the presence of BHs, it is reasonable to expect at low redshift that the stronger the accretion, the stronger the AGN feedback, thus the stronger the quenching of star formation. This should result in an overall reduced \ssfr, a behaviour that is indeed found when comparing the \ssfr iso-contours between the \hagn and \hnoagn simulations. Interestingly, Figures~\ref{fig:3D_ssfr_long} - \ref{fig:3D_curv_ssfr_cen_sat}
and Figure~\ref{fig:3D_curv_redshift_ssfr}  also show that the shape of the \ssfr iso-contours is modified in the presence of AGN feedback such that, at the high mass end, galaxies with highest \ssfr seem to be off-set from the highest density nodes of the cosmic web (see also Appendix~\ref{sec:AGN_noAGN_ssfr} which quantifies the difference of \ssfr between \hagn and \hnoagn).
Satellites are much less impacted by AGN feedback than centrals, and their \ssfr  is  mostly affected by the environment of groups and clusters.

\subsection{Evidence for  other processes driving galaxy formation}
\label{sec:hidden_variable}

Closer inspection specifically shows that the iso-contours of \ssfr, \vsig (Figures~\ref{fig:3D_ssfr_long} and ~\ref{fig:framesaddle_vsigage_long}) on the one hand, and stellar mass (Figure~\ref{fig:framesaddle_mass_long_trans}) on the other differ from one another. This suggests that there may exist hidden processes driving galactic physics  (beyond mass and local density).

\begin{figure*}
\centering
\includegraphics[width=0.75\textwidth]{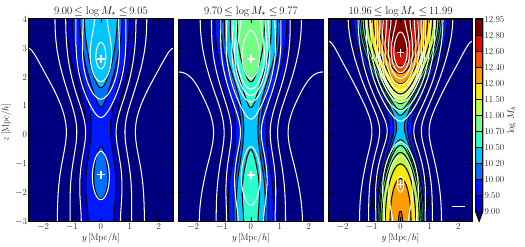}
\caption{Stellar mass weighted halo mass in the frame of the saddle at redshift zero for low ({\sl left}), intermediate ({\sl middle}) and high ({\sl right}) stellar mass bins, as labeled, in the longitudinal plane at the saddle. The shape of iso-contours does not change dramatically with stellar mass and not surprisingly, galaxies with highest stellar masses live in most massive haloes. Note that low values for halo mass result from the smoothing of mean values in sparsely occupied regions.
}
\label{fig:3D_curv_761_mhalo}
\end{figure*}

Let us attempt to quantify their nature.
Figure~\ref{fig:3D_curv_761_mhalo} displays the host's halo mass (resp. subhalo's mass for satellites) in the frame of the saddle, in the longitudinal cross section at redshift zero for different {\sl stellar} mass bins (see also Figure~\ref{fig:framesaddle_mass_long_trans}). Not surprisingly, galaxies with higher stellar mass are found to live in more massive dark matter haloes. These halo mass gradients are in agreement with Section~\ref{sec:theory}'s  theoretical prediction and reflect what was already seen for the stellar mass gradients of the entire galaxy population, 
i.e. saddle points represent maxima of the halo mass in the direction perpendicular to the filament, while they are minima in the direction along the filament towards nodes independently of stellar mass.
Note that in a given stellar mass bin, halo mass increases towards filaments and nodes, i.e. the $M_\star/M_h$ ratio is decreasing along those directions.
Strikingly, there is little change in the shape of these halo mass gradients  when varying stellar mass. This is strongly indicative that stellar mass is at first order only a function of
dark matter mass (at a given position within the cosmic web)\footnote{A tight correlation between the stellar and halo mass of galaxies in the current framework of galaxy formation \citep[][]{rees77,
fall80} is expected based on abundance matching \citep[e.g.][]{Conroy2009,mosteretal13,RodriguezPuebla2017} and confirmed with more direct measurements using e.g. satellite kinematics \citep[e.g.][]{vandenBosch2004,More2009} or weak lensing \citep[e.g.][]{Moster2010,Han2015,vanUitert2016}.}.
This is in sharp contrast with Figure~\ref{fig:3D_ssfr_long} (resp. Figure~\ref{fig:framesaddle_vsigage_long}), which shows that the \ssfr (resp. \vsig) contours do vary significantly across stellar mass bins
{\sl and} have also distinct shapes compared to  Figure~\ref{fig:3D_curv_761_mhalo}. 

\begin{figure*}
\centering
\includegraphics[width=0.75\textwidth]{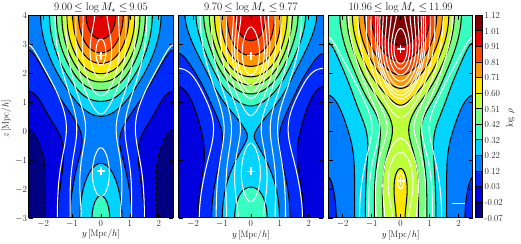}
\caption{Stellar mass weighted density in the frame of the saddle at redshift zero for low ({\sl left}), intermediate ({\sl middle}) and high ({\sl right}) stellar mass bins, as labeled, in the longitudinal plane at the saddle. As for halo mass (see Figure~\ref{fig:3D_curv_761_mhalo}), the shape of iso-contours does not change dramatically with stellar mass and not surprisingly, galaxies with highest stellar masses live in densest regions.
}
\label{fig:3D_curv_761_dens}
\end{figure*}

Besides halo mass, density is another obvious candidate for a variable that could drive the observed \ssfr (resp. \vsig) distributions in the frame of the saddle.
Figure~\ref{fig:3D_curv_761_dens} shows the density in the frame of the saddle, in the longitudinal cross section at redshift zero for different {\sl stellar} mass bins (see also Figure~\ref{fig:framesaddle_mass_long_trans}). This density is computed  on the scale of 0.8 Mpc/$h$, at which the skeleton was defined (and where the corresponding level of anisotropy was defined). 
Not surprisingly, galaxies with higher stellar mass are found to live in denser regions. 
These maps are again in agreement with Section~\ref{sec:theory}'s  theoretical prediction and are qualitatively similar to what was already seen for the halo mass gradients, i.e. saddle points represent maxima of the density in the direction perpendicular to the filament, while they are minima in the direction along the filament towards nodes independently of stellar mass. As  for halo mass, there is little change in the shape of  these maps versus stellar mass. 
This in turn may indicate that there exist other position-dependent   variables which impact \ssfr (resp. \vsig). 

\begin{figure*}
\centering
\includegraphics[width=0.95\textwidth]{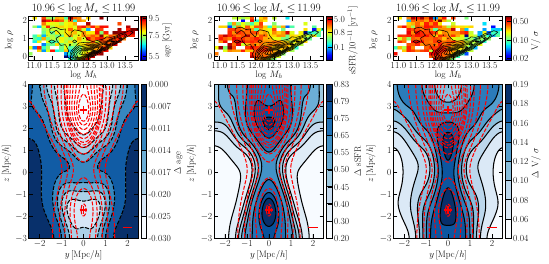}
\caption{Age, \ssfr and \vsig residuals, from {\sl left} to {\sl right}, having removed the mean stellar mass, halo mass and density effects, respectively by binning and considering the median mapping (see text for details), in terms of fraction of the median values in the frame of the saddle at redshift zero for high stellar mass bin, in the longitudinal plane at the saddle. The red dashed contours and the red crosses correspond to the galaxy number counts and the peaks in galactic density on axis, respectively and the red horizontal lines represent the smoothing length used in the analysis. The sub-panels on the top show the distributions of the three parameters, age, \ssfr and \vsig (in colours), as a function of halo mass, $M_{h}$ and local density $\rho$ (computed on the scale of 0.8 Mpc/$h$, at which the skeleton and the corresponding level of anisotropy were defined), respectively, with number counts overplotted in black. Interestingly, the residuals for \ssfr and \vsig display an excess at finite distance between the saddle and the nodes, which points towards the expected loci of maximum spin up and limited AGN quenching. Conversely, the age residuals are very small ($\lesssim 3 \%$) relative to the values of residuals obtained for \ssfr and \vsig, consistently with the observation that the age, halo mass and local density gradients show many similarities (compare bottom panel of Figures~\ref{fig:framesaddle_vsigage_long}, Figure~\ref{fig:3D_curv_761_mhalo} and Figure~\ref{fig:3D_curv_761_dens}, respectively). 
}
\label{fig:3D_curv_761_delta}
\end{figure*}

Let us attempt to quantify this effect by calibrating from the full 
 simulation the mapping $\hat \ssfr(M_{\rm h},\rho)$, defined as the median \ssfr at given $M_{\rm h}$ and local density $\rho$ (and $M_\star$ given that the mapping is defined in a given stellar mass bin), where $\rho$ is computed on the scale of 0.8 Mpc/$h$.
To do this, the median \ssfr is computed in bins of $\log M_{\rm h}$ and $\log \rho$ constructed adaptatively such that each of 10 equipopulated bins of $M_{\rm h}$ is further divided in 8 equipopulated bins of $\log \rho$, in a given stellar mass bin. 
This median relation is then used in a 2D interpolation to obtain a relation $\hat \ssfr(M_{\rm h}, \rho)$ that can be applied to each galaxy (see Appendix~\ref{sec:stat} for details). 
Should the physical process driving star formation only depend on mean density and mass\footnote{
Note  that we cannot  rule out that position-dependent shape of the PDF of the distribution of \ssfr, halo mass and density accounts for some residuals,
as one would not expect the averaging and the mapping to fully commute, see Appendix~\ref{sec:stat}. }, this operation would reproduce exactly
Figure~\ref{fig:3D_ssfr_long}. 
What is found instead is that at given stellar mass, there is a clear position-dependent discrepancy between the two, as shown in Figure~\ref{fig:3D_curv_761_delta}\footnote{We also computed maps of the density smoothed on 2 and 3 Mpc/$h$. This had little impact on the equivalent of Figure~\ref{fig:3D_curv_761_delta} while significant residuals are found  at the lower stellar mass bin, as expected since the smaller the mass the smaller the scale and the smaller the correlation with the field smoothed on larger (fixed) scale. This is consistent with the findings of \cite{Kraljic2018}.}. 
This figure displays the difference of the mean \ssfr measured at the given position, and the mean \ssfr estimated using the above-defined mapping, 
in highest stellar mass bin, normalised by the median \ssfr (computed over the whole saddle region).
This discrepancy  is indicative that the impact of the saddle accounts for at least a fraction of the dispersion from the median $\ssfr-M_{\rm h}-\rho$ relation (middle and right panels), either because of the imposed local tides and/or because of the scatter in density imposed by this saddle (which might also be position-dependent).
Interestingly, when the same transformation is applied to galactic age, no significant residuals are found (left-hand panel). This suggests that mean stellar mass and
age, which are integrated quantities, do not seem to be very sensitive to anything but mean dark halo mass and mean density.
Appendix~\ref{sec:stat} discusses in more details how to statistically disentangle mass, density and tidal effects.

\subsection{Is spin advection one of the residual processes?} 
\label{sec:spin}

\begin{figure}
\centering
\includegraphics[width=0.9\columnwidth]{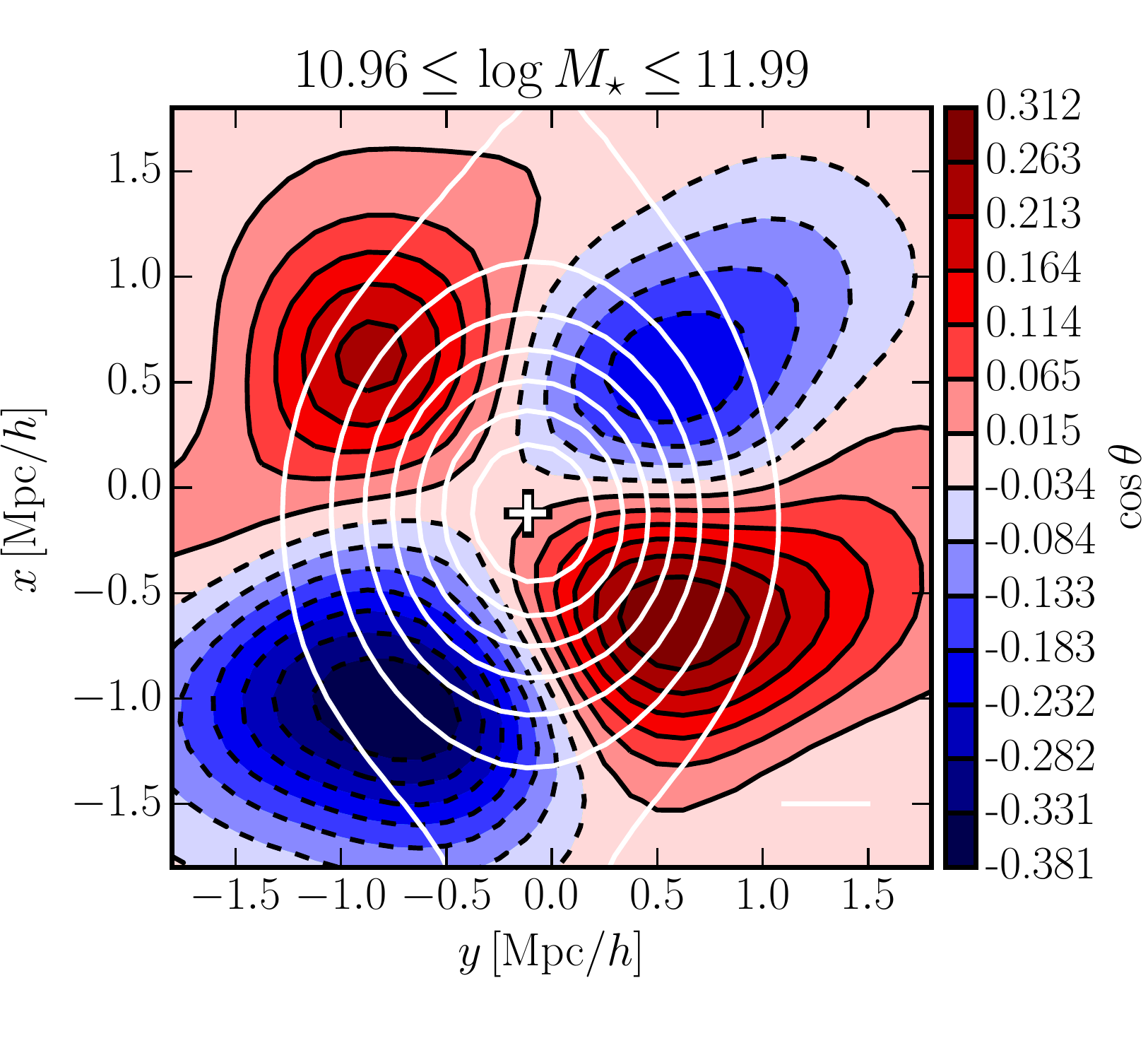}\\
\includegraphics[width=0.9\columnwidth]{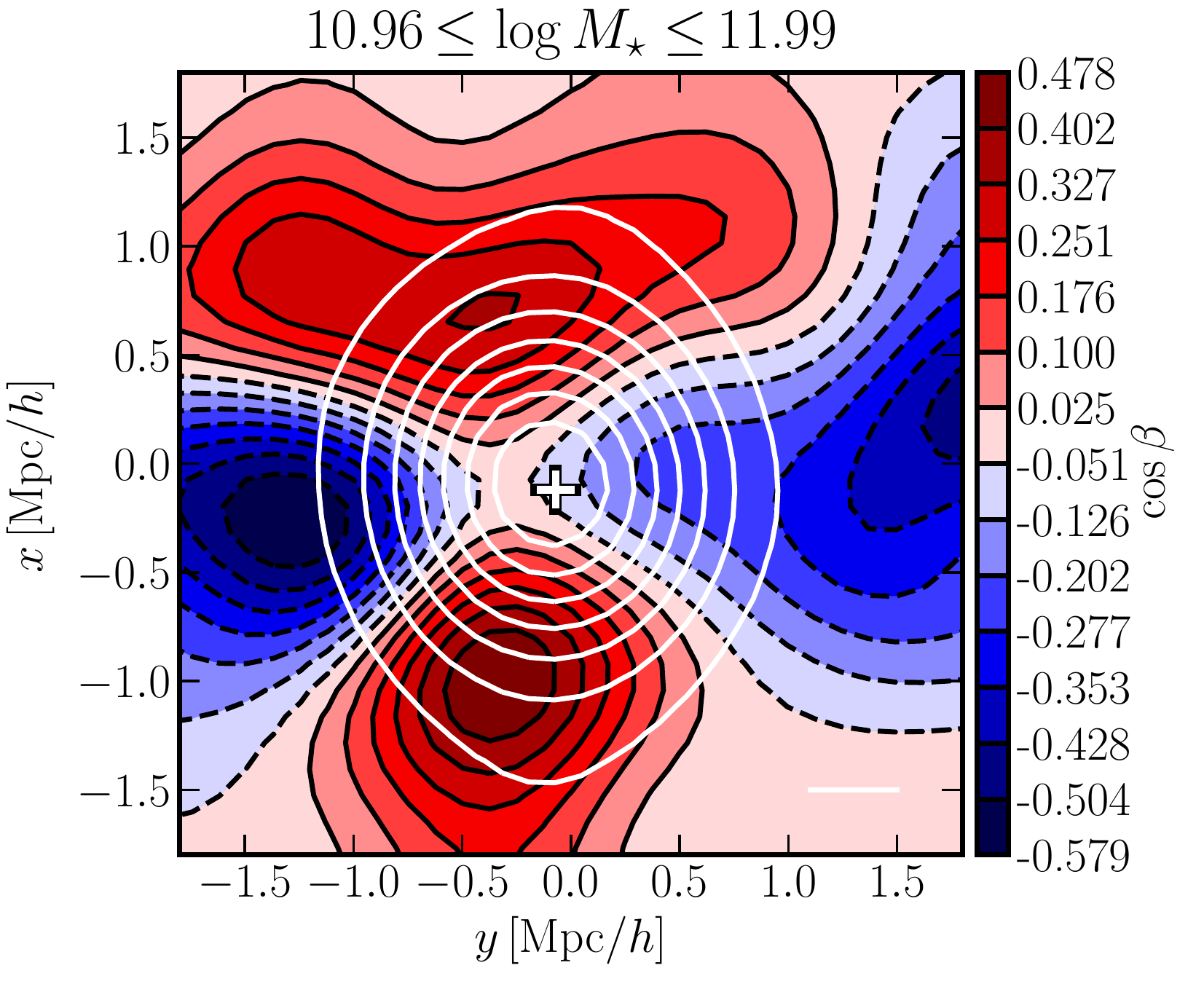}
\caption{Stellar mass-weighted cosine of angle between the spin of the galaxy and the direction of the filament ({\sl top panel}) and between
the vorticity of the gas at the position of the galaxy and the direction of the filament ({\sl bottom panel}) for highest stellar mass bin at redshift zero. Transverse cross-sections comprise galaxies with $z$-coordinate between 1 and 2.5 Mpc/$h$ for the spin and between 0.5 and 1.8 Mpc/$h$ for the vorticity, where the signal is most significant.
The vorticity is computed on the gas distribution at the resolution and smoothing length used to define the skeleton, and interpolated at the position of galaxies.
The map is normalised so that the integrated amplitude in each quadrant is preserved while smoothing. 
The white contours and the white crosses correspond to the galaxy number counts and the peaks in galactic density, respectively. 
Note the quadrupolar (point-reflection symmetric) structure of the spin's orientation distribution in the vicinity of the filament axis, in qualitative agreement with the prediction of  \protect\cite{Codis2015a} for dark matter.
The distribution of the vorticity of the gas is also in qualitative agreement with this prediction and  with the measurement of  \protect\cite{laigle2015} (their Appendix A), which focused on cooling runs (without star formation). The tilt in the plane of symmetry of the vorticity map is likely to be  driven by shot noise.
}
\label{fig:spin_highM}

\end{figure}

In closing, let us speculate on the nature of the physical process  which may be responsible for the residual scatter -- having removed  some of the effect of mean mass and local density, while relying on our
saddle-centred stacks to identify processes that may be driven by  anisotropy.
As already mentioned, the (radial) distance to the node-quenching from AGN feedback is an obvious candidate for the amplitude of the residual maps.
Nevertheless, it had long been known that  angular momentum stratification --    undoubtedly built from anisotropic tides -- is  a key underlying  property driving morphology of galaxies, which correlates  with their star formation efficiency.
Angular momentum acquisition is controlled by the large-scale tidal tensor, which  imprints its torque along the galaxy's  lightcone.
The induced tides  not only impact the assembly and accretion history of the host, but also the filamentary flow of cold gas connecting to the host,  hence its coherent gas supply.
It has recently been shown  \citep{welkeretal2015b}  following galaxies that the quadrupolar vorticity-rich large-scale filaments  are indeed the loci where low- and intermediate- mass galaxies steadily acquire angular momentum via quasi-polar cold gas accretion, with their angular momentum  aligned with the host filament~\citep[see Figure~\ref{fig:spin_highM} for the high mass bin which  has the most significant alignment signal at low redshift, and][]{laigle2015}: galaxies are expected to accrete more efficiently cold gas when their angular momentum is aligned with the preferential direction of the gas infall, \textit{i.e.} aligned with the filament \citep{Pichon2011,stewart2011}.
This has typical local kinematic signatures in terms of i) spin and ii) vorticity orientation as predicted by \citep{Codis2015a}, and as measured in \hagn w.r.t. the direction of its closest filament (Figure~\ref{fig:spin_highM}),
and iii) in terms of internal kinetic anisotropy in the velocity dispersion of dark halos \citep{Faltenbacher:2009fl}.
The \vsig of galaxies  increases  as they  drift along the filament without significant merger, as they align themselves to the saddle's tides (Figure~\ref{fig:spin_highM}).

\begin{figure}
\centering
\includegraphics[width=0.5\textwidth]{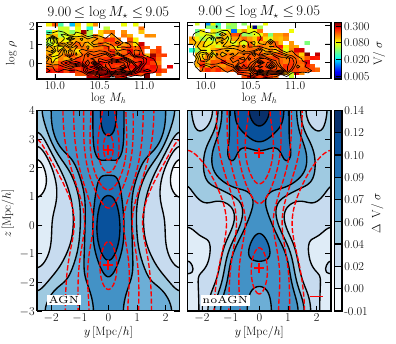}
\caption{\vsig residuals in \hagn ({\sl left panel}) and \hnoagn ({\sl right panel}), having removed the mean stellar mass, halo mass and density effects, respectively by binning and considering the median mapping (see text), in the lowest stellar mass bin at redshift zero. The red dashed contours and the red crosses correspond to the galaxy number counts and the peaks in galactic density on axis, respectively and the red horizontal lines represent the smoothing length used in the analysis.
The two maps are qualitatively similar, as expected for this mass bin where AGN feedback should not have a strong impact, and nonetheless the amplitude of the map is $\sim$ 14\%,
concentrated along the filament's axis. 
}
\label{fig:3D_residuals_AGN}
\end{figure}

The  efficiency of star formation, as traced by  sSFR,  also depends  on the infalling rate and impact parameter of the cold gas in the circum-galactic medium.
Hence one  also expects star formation efficiency to be strongest wherever the alignment is tightest. The locus of this induced excess of star formation and/or \vsig should therefore have measurable signatures in observations when quantified in the metric of the filament  
\citep[as discussed e.g. in][Eqn. 40, in terms of loci of maximum  cold gas advection at some finite distance from the saddle along the filament]{Codis2015a}.  
There is a hint of such excess in the residuals shown  in Figure~\ref{fig:3D_curv_761_delta} in terms of both \ssfr and \vsig  (which  should co-evolve). While quenching is  also  playing a significant  position-dependent role for the high mass population, its impact on the lower mass galaxies
will be less significant. Figure~\ref{fig:3D_residuals_AGN} shows indeed that for the lower mass bins, the residual maps peak significantly on axis,
which supports the idea that the efficiency  of angular momentum advection is a relevant process.
This  is worth emphasising, given the above-given theoretical prejudices based on following galaxies in the flow \citep{welkeretal2015b}, and on the orientation of galaxies traced by their 
spin's orientation distribution in the vicinity of the filament axis, predicted to exhibit a point-reflection symmetric structure \citep[][]{Codis2015a}
as  measured in Figure~\ref{fig:spin_highM}.
While this discussion is more speculative, recall  in any case that most  properties of the galactic population measured within the frame of saddles presented in the previous section -- including redshift evolution and filament thickening/thinning -- can be understood when accounting for their cosmic advection with the bulk flow along and transverse to the filament. 
The present study  clearly highlighted that 
an improved model for galaxy properties should also explicitly integrate the diversity of the  topology of the large environment on multiple scales \citep[following, e.g. ][]{Hanami2001}
and quantify the impact of its anisotropy on galactic mass assembly history, and more generally on the kinematic history of galaxies. The details of how the kinematics impact star formation remains to be understood.
The vorticity-rich kinematics of the large scale flow is neither strictly coherent  nor fully turbulent. Does the offset of merger and accretion rate imposed by the large scale turbulent  flow  explain the residual environment dependence in observed physical properties \citep{CWD2016}, or is the helicity of gas inflow within filaments prevalent in feeding galactic discs coherently \citep{pichonetal11}?

\section{Conclusions}
\label{section:conclusion}

This paper investigated  the properties of virtual galaxies in the neighbourhood of filament-type saddle points of the cosmic web. These properties were measured within the frame set by the principal axes of the saddle in the \hagn simulation. The impact of AGN feedback  was assessed by comparing to  results obtained in the \hnoagn simulation.
The principal findings are the following:
\begin{itemize}
\item The iso-contours of the galactic number density, mass, \ssfr, \vsig and age in the saddle's frame display a clear alignment with the filament axis and stronger gradients
perpendicular to the filaments,  quantifying the impact of the cosmic web in shaping galaxies.  
\item High mass galaxies are more clustered around filaments and within filaments around nodes compared to their low mass counterparts.  As expected, the filament's width of the whole galaxy population  grows with cosmic time (as it becomes dominated by less rare galaxies). Conversely, at fixed mass, it decreases with cosmic time at the saddle.
\item In addition to reducing the overall \ssfr of galaxies, AGN feedback  also impacts  the shape of the \ssfr iso-contours, in particular for high mass galaxies and in the vicinity of the nodes of the cosmic web. AGN feedback quenches centrals more efficiently than satellites. Satellite strangulation seems to occur within the filaments and nodes of the cosmic web.
\item While the dominant effect of the cosmic web on galaxy formation seems to be captured by the distance to cosmic nodes, the full three dimensional geometry of the web, in particular its saddle points, provides a natural oriented frame for stacking galaxies, showing significant effects of the environment beyond solely the distance to nodes. Hence, galaxies do retain a memory of the large-scale cosmic flows from which they emerged.
 \item The redshift evolution of the galactic counts  and the age distribution of galaxies are consistent with a drift of the population towards
the filaments and along them (see Appendix~\ref{sec:vel}). The cosmic evolution of the sSFR reflects both this drift and the triggering of quenching as centrals become massive enough to trigger AGN feedback near the peaks of the cosmic web. The geometry of the stacks and their cosmic evolution compare favourably to expectations for constrained Gaussian random fields in the weakly non-linear regime.
\item  The maps of $V/\sigma$  and \ssfr (and their residuals) are consistent with the role played by feedback {\sl and} angular momentum in shaping galaxies, beyond that played by mass and density, and its connection with the geometry of the cosmic web, as described  by \cite{Codis2015a} and  \cite{laigle2015} (in a Lagrangian  and an  Eulerian framework, respectively). The point-reflection symmetric distribution of the orientation of the spin of galaxies and vorticity of the gas presented in this paper is also
in agreement with this picture.
\item  At high mass and low redshift, AGN feedback coupled with advection of galaxies along filaments induces some level of anisotropy in the distribution of galaxy properties (\ssfr, \vsig, age) which is partially degenerate with the effect of how angular momentum of galaxies is acquired from the large-scale vorticity of the anisotropic environment.
\item  While \ssfr responds to the saddle frame over and above what is expected from halo mass and local density,  other indicators such as stellar age do not. 
\end{itemize}
Overall, all distributions are consistent with the geometry of  the flow in the vicinity of saddles, 
including quenching by AGN feedback, strangulation of satellites near the nodes, and possibly time delays induced by asymmetric tides on local {\sl and} intermediate scales.
They complement the findings of \cite{Kraljic2018}, which also showed that galaxy properties occupy more than a two dimensional manifold (in physical parameter space such as age, sSFR, $V/\sigma$ etc.), but at the expense of not resolving the 3D distribution of fields in the frame of the saddle, which was the adopted strategy here\footnote{Alternatively,
one could stack in the theoretically motivated \citep{Musso2018}  `natural' 2D frame of the saddle  using radius, $r$,  and `angle', ${\cal Q}= \mathbf{r}\cdot {\bar{ \mathbf{H}}} \cdot \mathbf{r}/r^2 $, where $\bar{ \mathbf{H}}$ is the tidal tensor.}.
This strategy allows us to {\sl suggest} that one extra degree of freedom is the angular momentum acquired from the anisotropy of the cosmic web.

The signal-to-noise ratio in the counts is in the current analysis limited by the number of galaxies in the simulated box and by the choice of sampling the population in 3D.
In order to e.g. probe the transverse asymmetry of saddles (reflecting the relative depth and distance to neighbouring voids and wall-saddles), the present study could be followed up using simulations with better statistics so that the counts may be orientated w.r.t. the connecting walls and voids.
A larger sample would also allow us to quantify the effect of non-linearities when constructing residual maps, as discussed in Appendix~\ref{sec:stat}.
It would also be of interest to stack  observationally  measurable quantities such as colour  or metallicity.
These predictions could then be directly compared to observations from upcoming spectroscopic surveys such
as 4MOST \citep{4MOST}, DESI \citep{DESI}, PFS \citep{PFS}, MSE \citep{2016arXiv160600043M},
integral field spectroscopy such as  MANGA \citep{Manga2015}, SAMI \citep{sami2012}, Hector \citep{Hector2015}
  or in projection using photometric redshifts  with  DES \citep{DES2016}, Euclid \citep{Euclid}, WFIRST \citep{WFIRST}, LSST \citep{LSST}, KiDs \citep{KIDS},  following the pioneer work of  \cite{Laigle2018} in the COSMOS field.
Connecting the present findings with work on spin orientation \citep{Codis2015a}
 in the frame of the saddle may also prove useful
to mitigate the effect of  intrinsic alignment \citep[e.g.][]{Joachimi2011,Chisari2015}.
Investigating the distribution and survival of filaments on much smaller scales  as they enter dark haloes is also of interest and will be the topic of future work (Darragh Ford et al. {\sl in prep.}).

\section*{Acknowledgments}
{\sl We thank Romeel Dav\'e and Shadab Alam for stimulating discussions, and Elisa Chisari, Dmitry Pogosyan and Raphael Gavazzi and the anonymous referee for their comments which helped to improve this work.
KK thanks Joanne Cohn for fruitful discussions and helpful comments.
This work was granted access to the HPC resources of CINES (Jade) under the allocation 2013047012 and c2014047012 made by GENCI.
This research is part of the Spin(e) (ANR-13-BS05-0005, \href{http://cosmicorigin.org}{http://cosmicorigin.org}), Horizon-UK projects and  ERC grant 670193.
SC thanks the Merac Fondation for funding.
We  warmly thank S.~Rouberol for running  the {Horizon} cluster on which the simulations were  post-processed.
CP thanks Churchill college and the Royal Observatory Edinburgh for hospitality while this work was resp. initiated and completed, and the \href{https://www.supa.ac.uk}{SUPA} distinguished visitor programme for funding.
CP thanks the community of \href{http://mathematica.stackexchange.com}{mathematica.stackexchange} for help.
}

\bibliographystyle{mnras}
\bibliography{author}

\appendix

\section{Validation} \label{sec:validation}

\begin{figure}
\centering
\includegraphics[width=0.45\textwidth]{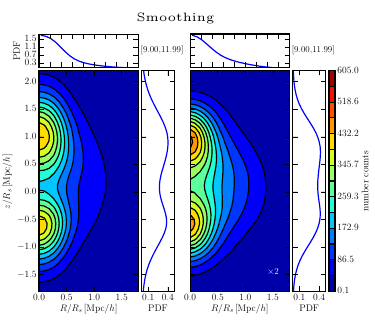}
\caption{
 Galaxy number counts in the frame of the saddle (curvilinear coordinates) for all masses in the range $10^{9.0}$ to $10^{12.0} M_\odot$ at redshift zero ({\sl left}) compared to the smoothing twice as big ({\sl right}).
 Note that the $R$ and $z$ axes have been rescaled by the smoothing length. Similarity of the contours suggest that the measured distributions are relatively insensitive to the choice of the smoothing length.
}
\label{fig:2D_curv_validation}
\end{figure}

Let us briefly study how the measured distributions presented in the main text are impacted by the smoothing length of the gas density distribution and the type of tracer used to extract the skeleton.
Results are presented in the frame of the saddle using the curvilinear coordinates (see Section~\ref{section:2Dgalaxies}), but qualitatively similar conclusions are obtained for 3D distributions.
Figure~\ref{fig:2D_curv_validation} shows the galaxy number counts for the entire galaxy population with masses in the range $10^{9.0}$ to $10^{12.0} M_\odot$ at redshift zero, using, after rescaling,  the  same smoothing length as in the main text ({\sl left}) and twice as big ({\sl right}). Similarity of these iso-contours suggests that as expected, the measured distributions are relatively insensitive to the level of smoothing applied.

\begin{figure}
\centering
\includegraphics[width=0.45\textwidth]{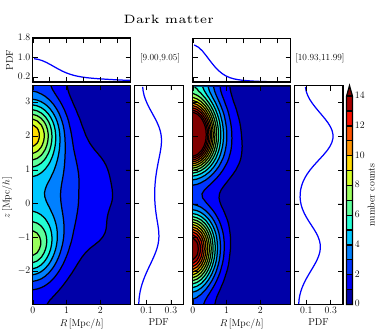}
\caption{
Galaxy number counts in the frame of the saddle (curvilinear coordinates) in low ({\sl left}) and high ({\sl right}) stellar mass bins at redshift zero, using the dark matter as a skeleton tracer. Similarity between these contours and those obtained using gas (see Figure~\ref{fig:2D_curv_761_nb_allM_mass}) suggest that the measured distributions are relatively insensitive to the choice of the tracer used to construct the skeleton.
}
\label{fig:2D_curv_validation2}
\end{figure}

Figure~\ref{fig:2D_curv_validation2} shows the galaxy number counts in low ({\sl left}) and high ({\sl right}) stellar mass bins at redshift zero, using the dark matter particles as a tracer of the cosmic web. These iso-contours that should be compared with left and right panels of Figure~\ref{fig:2D_curv_761_nb_allM_mass}, suggest again only a weak dependence of results on the choice of the tracer (i.e. gas or dark matter).
Note nonetheless that the skeleton built directly from galaxies using persistence is significantly different, as it becomes multi-scale in nature. The corresponding complication is beyond the scope of this paper and will be explored elsewhere.

\section{Filaments' length and width}
\label{sec:filaments}

\begin{figure}
\centering
\includegraphics[width=0.8\columnwidth]{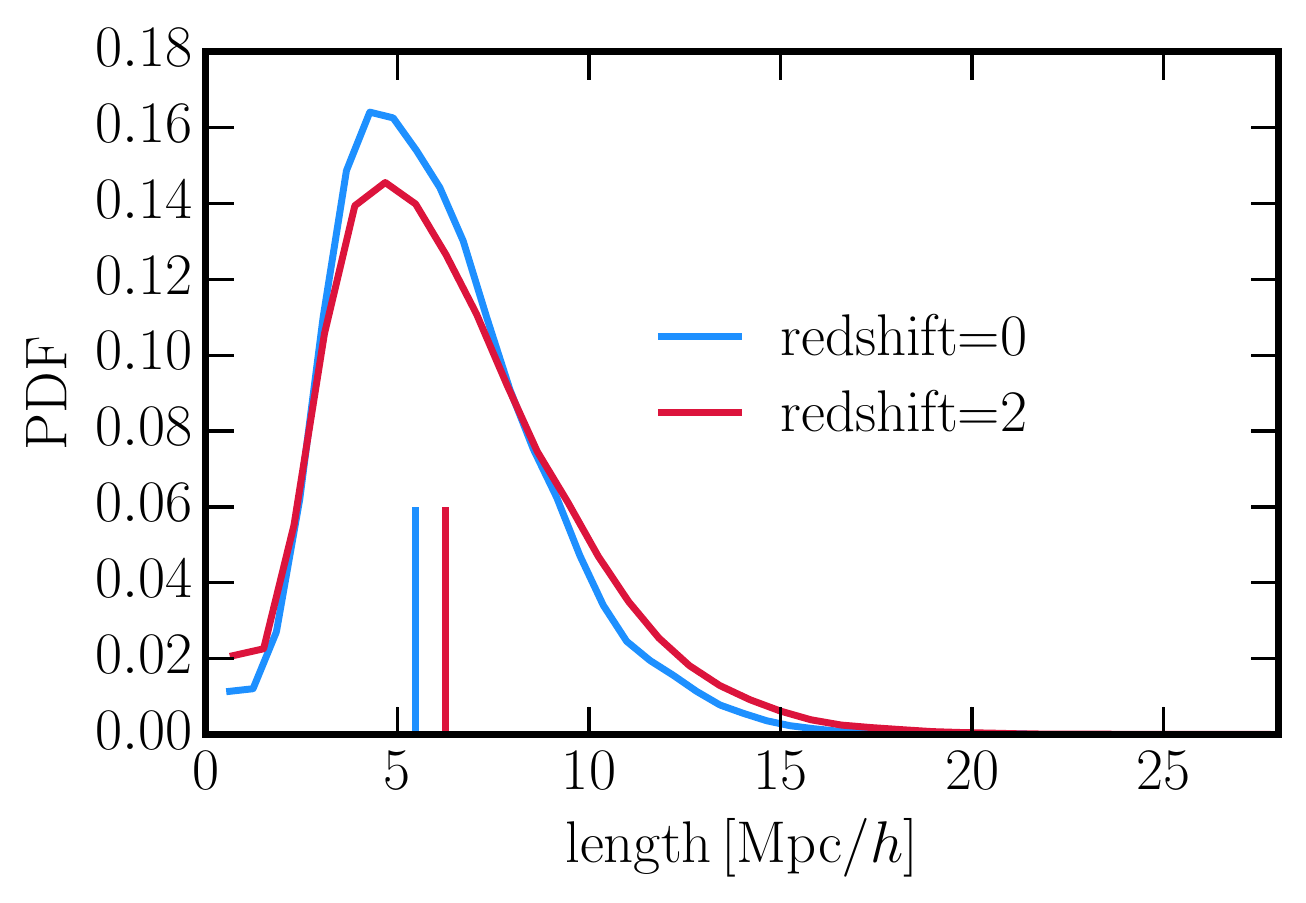}
\caption{Probability distribution function of the length of filaments at redshift two (red) and redshift zero (blue). The vertical lines correspond to the medians of distributions.
}
\label{fig:fil_len}
\end{figure}

\begin{figure}
\centering
\includegraphics[width=0.8\columnwidth]{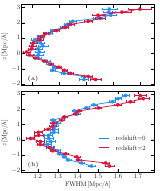}
\caption{Thickness of the filaments, defined as the FWHM of the Gaussian fit of the transverse galaxy number counts profiles as in Figure~\ref{fig:filament_thickness}.
When considering the entire population of galaxies, the cross sections of filaments in the vicinity of the saddle point (at $z=0$ Mpc/$h$) grow with time, while in the vicinity of nodes ($z \sim 2.5$ Mpc/$h$ and $z \sim -1.5$ Mpc/$h$ for highest and lowest density nodes, respectively), they get thinner.
}
\label{fig:fil_thick}
\end{figure}

Figure~\ref{fig:fil_len} shows the probability distribution of the length of filaments at redshift two and zero. The length of filaments decreases with time, in agreement with the expected evolution of matter distribution in the $\Lambda$CDM universe with accelerated expansion at redshift $\lesssim 1$  and as measured by \cite{Sousbie2008} for the dark matter. As universe expands, more low mass objects form leading to the formation of filaments on smaller scales that eventually merge together while longer filaments are stretched. Because larger scale filaments are less numerous than filaments on small scales, the net result is a shift of the median length towards lower values at lower redshift.

Figure~\ref{fig:fil_thick} shows the thickness of the filaments as a function of the position in the direction along the filament and is complementary to Figure~\ref{fig:filament_thickness} in that it extends to the vicinity of the nodes.
Regions near the nodes (in both upper and lower directions from the saddle, corresponding to the nodes of highest and lowest density, respectively) are getting thinner with time.

\section{Azimuthally averaged  sections}
\label{sec:azimuthal}

\begin{figure*}
\centering
\includegraphics[width=0.9\textwidth]{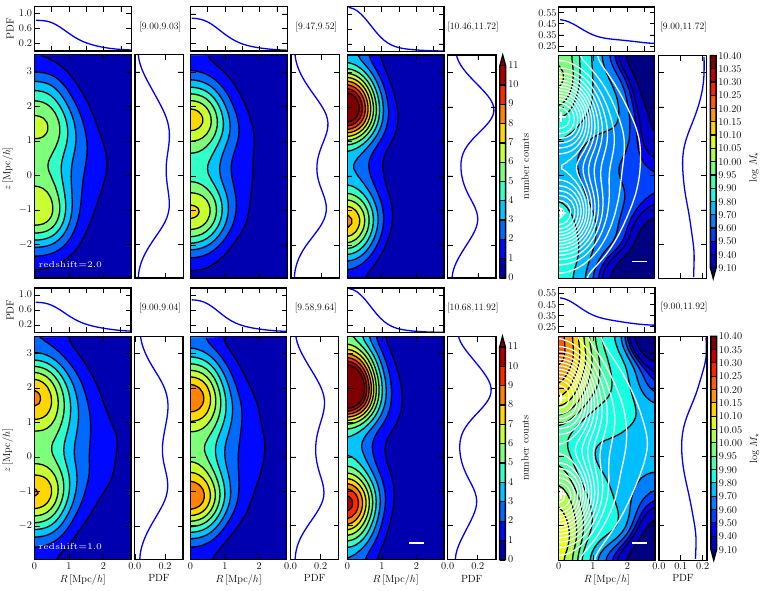}
\caption{ {\sl Left and middle panels:} Redshift evolution of the galaxy number counts in low ({\sl left column}), intermediate ({\sl middle column}) and high ({\sl right column}) stellar mass bins, in the frame of the saddle (curvilinear coordinates) at redshift two ({\sl top row}) and one ({\sl bottom row}), respectively. The white horizontal lines represent the smoothing length used in the analysis.
 {\sl Rightmost panels:} Redshift evolution of mean stellar mass iso-contours in the frame of the saddle (curvilinear coordinates) for the entire galaxy population at redshift two ({\sl top row}) and one ({\sl bottom row}), respectively. The white curves correspond to the contours of the galaxy number counts, while the white crosses represent the peaks in galactic number density on axis. Note how galaxies become more clustered towards filaments and nodes as they grow in mass with decreasing redshift, consistently with the global flow of matter within the cosmic web and in agreement with results considering the 3D distributions.
}
\label{fig:2D_1D_curv_redshift_nb_mass}
\end{figure*}

All distributions presented in Section~\ref{section:3Dgalaxies} -- considering the stacks in 3D, and Section~\ref{section:redshift} presenting their redshift evolution -- are in qualitative agreement with azimuthally averaged maps in 2D, adopting curvilinear coordinates as in Section~\ref{section:2Dgalaxies}.
Let us here focus on redshift evolution alone.
Figure~\ref{fig:2D_1D_curv_redshift_nb_mass} shows the galaxy number counts and mean stellar mass in the frame of the saddle using curvilinear coordinates, at redshifts two and one, complementing Figure~\ref{fig:2D_curv_761_nb_allM_mass}.
The redshift evolution of both number counts and mean stellar mass is in qualitative agreement with the results obtained when considering stacks in 3D (see Section~\ref{section:redshift}) and consistent
with the global flow of matter towards the filament first and along them afterwards \citep[see][for the dark matter flow]{Sousbie2008}.

\section{AGN quenching efficiency}
\label{sec:AGN_noAGN_ssfr}

\begin{figure}
\centering
\includegraphics[width=0.7\columnwidth]{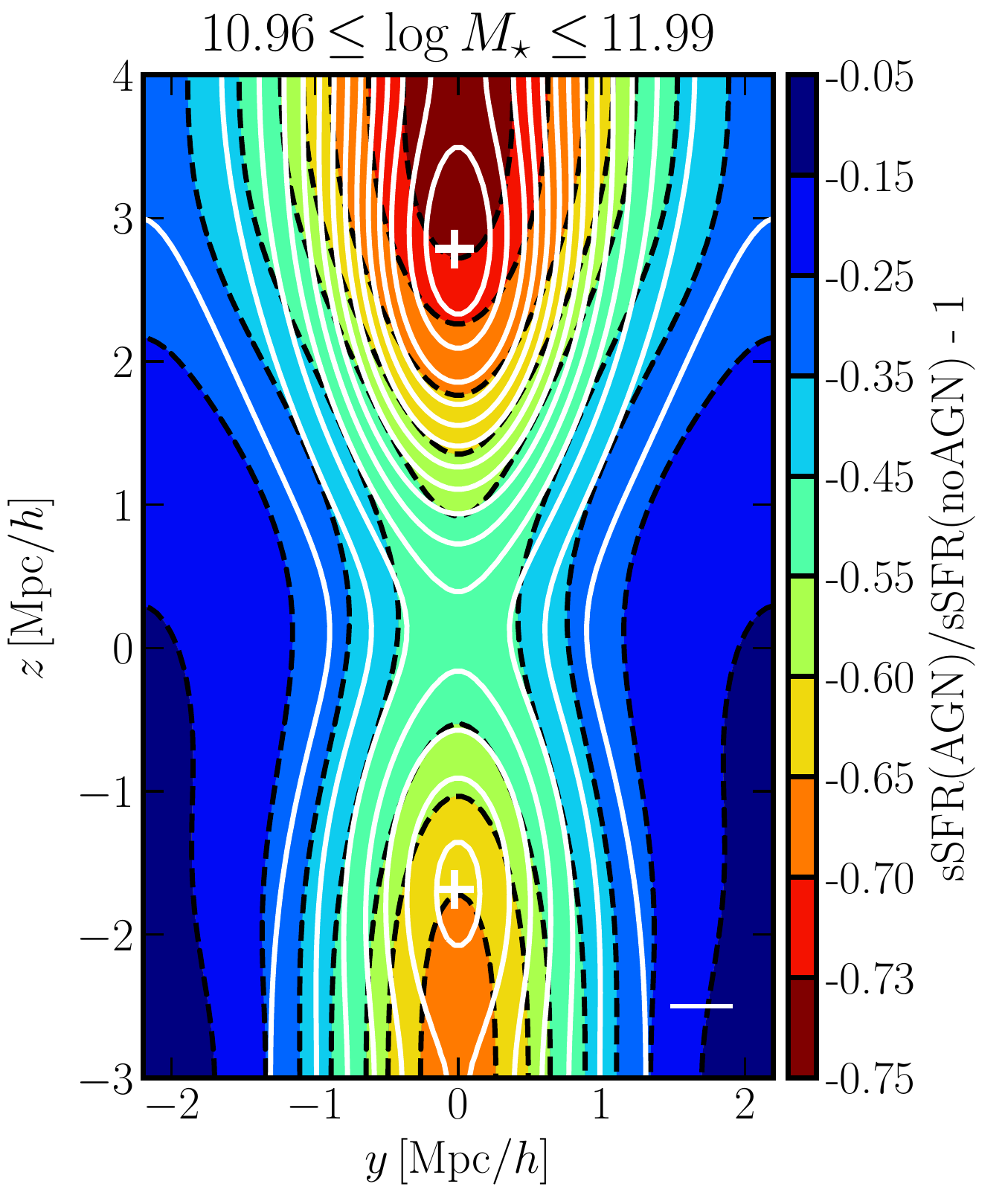}
\caption{\ssfr ratio in the \hagn, \ssfr(AGN), and \hnoagn, \ssfr(noAGN), simulations at redshift zero for highest stellar mass bin. Note that the highest impact of AGN feedback on \ssfr of galaxies is, as expected, in the vicinity of the densest node. White contours represent galaxy number counts in the \hagn simulation.
}
\label{fig:AGN_noAGN_ssfr}
\end{figure}

Figure~\ref{fig:AGN_noAGN_ssfr} shows the normalised difference of \ssfr in the \hagn and \hnoagn simulations at redshift zero for highest stellar mass bin. This quantity allows to quantify where the quenching is most efficient. As expected, highest reduction of the \ssfr is in the vicinity of the densest node.

\section{Theoretical  predictions}
\label{section:theory}
\begin{figure}
\centering
\includegraphics[width=\columnwidth]{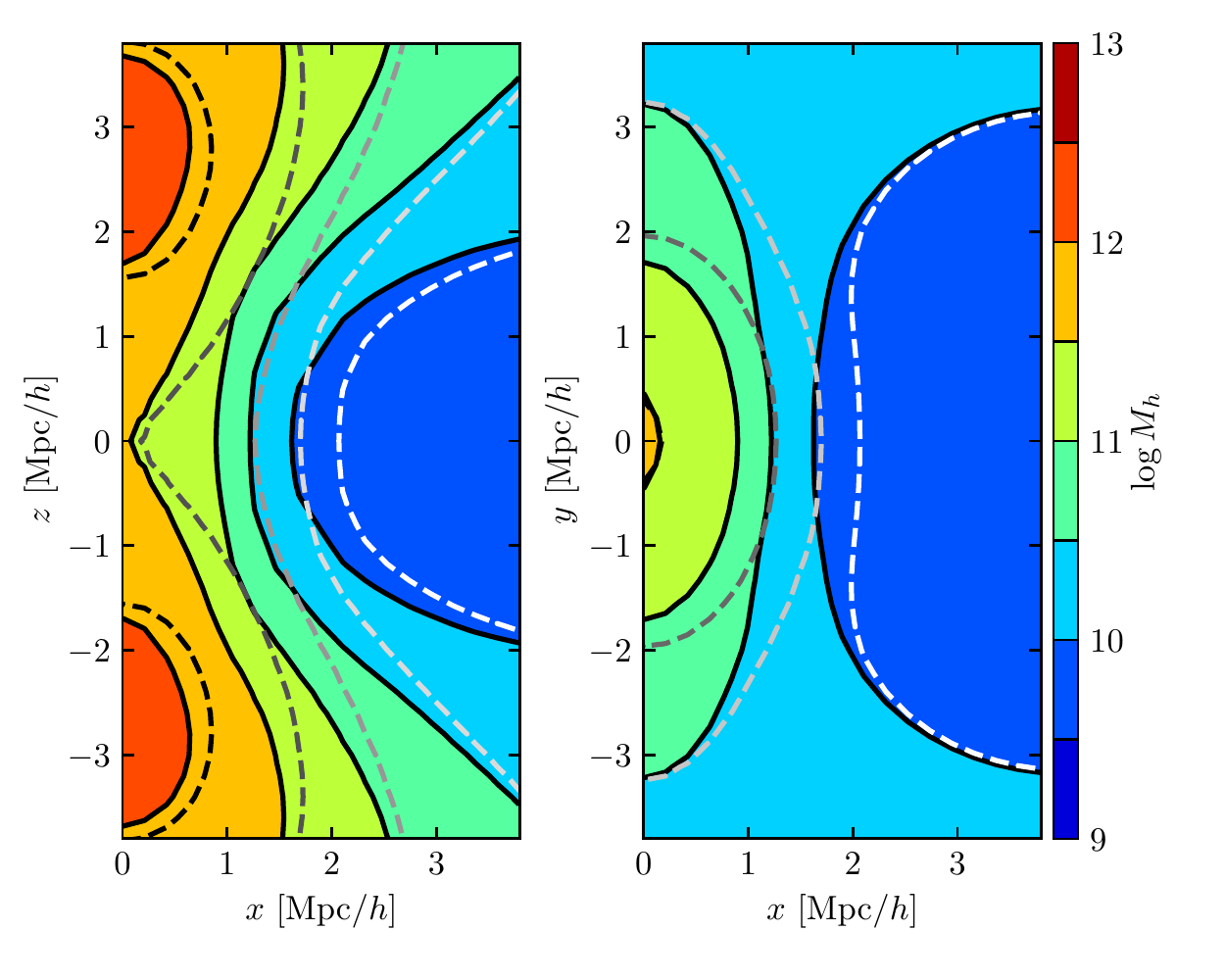}
\caption{{\sl Coloured contours:} predicted cross section of the halo mass density after a Zel'dovich boost for a fixed geometry of the saddle in the plane of the wall and filament (left panel) and perpendicular to the filament (right panel).
 {\sl Dashed contours:} the cross section before the boost. }
\label{fig:zeldoboost}
\end{figure}

\begin{figure}
\centering
\includegraphics[width=0.45\columnwidth]{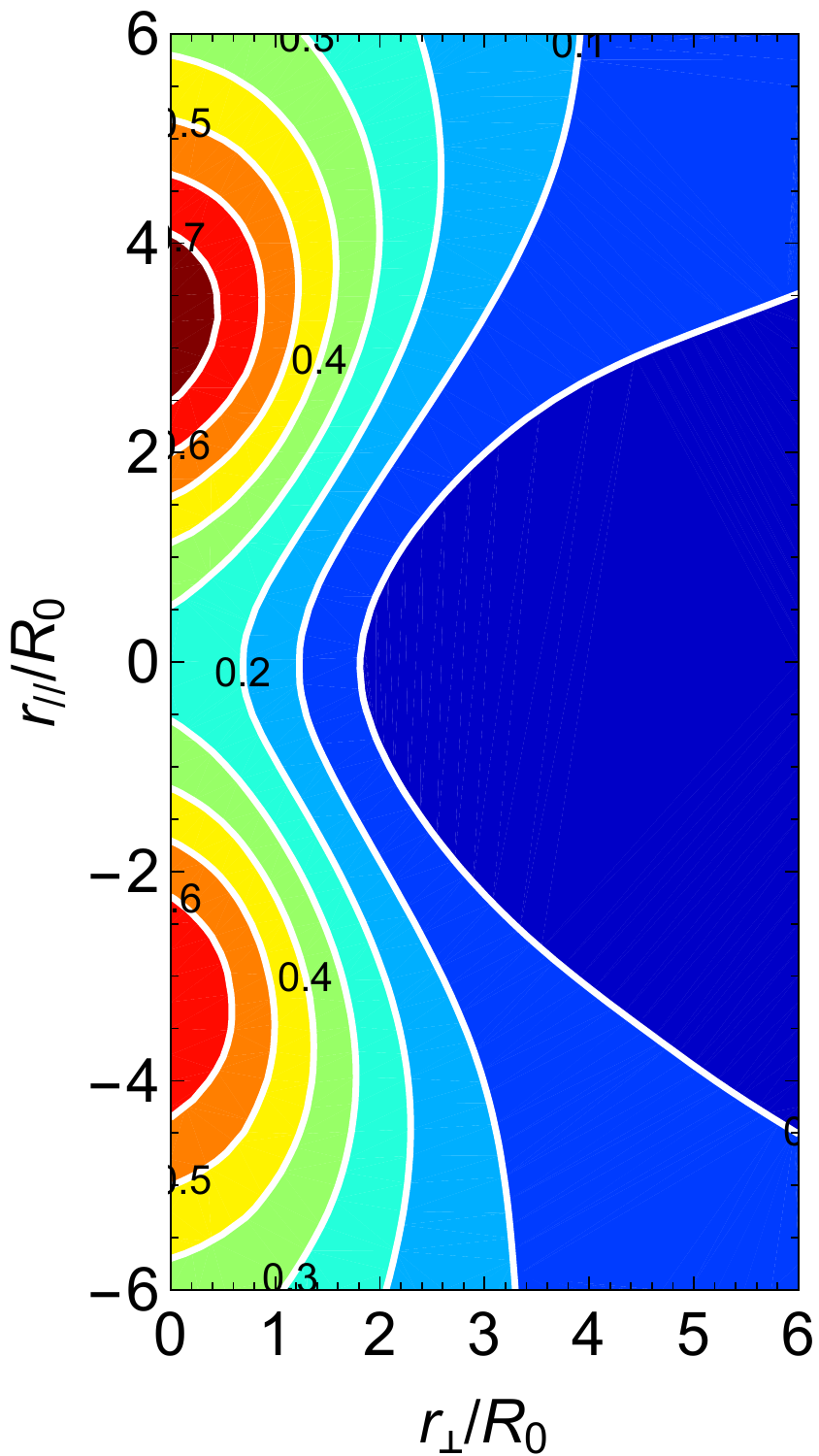}
\includegraphics[width=0.45\columnwidth]{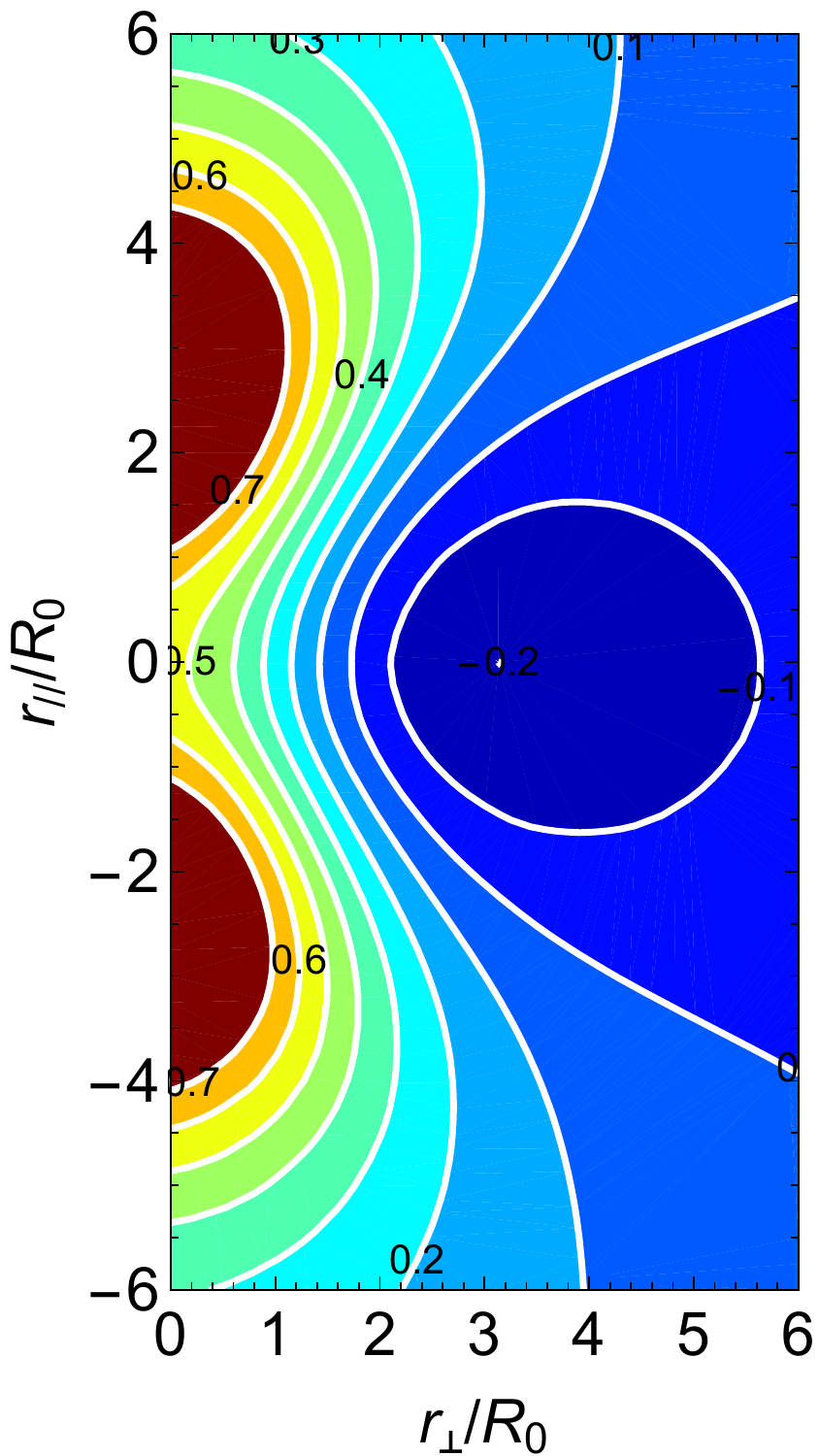}
\caption{ {\sl Left panel:} dark matter longitudinal cross section measurements in \hagn at redshift 0.  {\sl Right panel:} same quantity at redshift 2.
The corresponding prediction at high redshift is shown in Figure~\ref{fig:halo-mass-comparison} and the
agreement is fairly good.
}
\label{fig:mass-evolution}
\end{figure}

Let us briefly predict from first principles the expected shape of the matter and halo distribution in the vicinity of a saddle
and its cosmic evolution.

\subsection{Predictions for the mean constrained initial density field} \label{sec:GRF}

The initial density field in which the cosmic web develops being Gaussian, the theory of constrained Gaussian random field provides a natural framework
in which to compute the expectation of the matter distribution and typical halo mass within the frame set by the saddle point, as we do not expect the dynamics to be strongly non-linear on such scales.
An important ingredient here is therefore to impose a filament-type saddle point constraint.
Such a critical point form when the gradient of the density field is zero and is defined by its geometry, namely i) its height $\nu$ defined as the density contrast divided by its rms $\sigma_0=\sqrt{\left\langle \delta^2\right\rangle}$  and ii) its curvature by means of the three eigenvalues of the Hessian matrix of the density contrast rescaled again by their rms  $\sigma_2=\sqrt{\left\langle (\Delta\delta)^2\right\rangle}$. For a filament-type saddle point, $\lambda_{1}\geq0\geq\lambda_{2}\geq\lambda_{3}$.

The so-called peak theory \citep{Kac1943,Rice1945} then allows us to predict all statistical properties of critical points once the (supposedly Gaussian here) probability density function (PDF) of the field $\nu=\delta/\sigma_0$, its first $\nu_i=\delta_{,i}/\sigma_1$ and second derivatives $\nu_{ij}=\delta_{,ij}/\sigma_2$ is known. The saddle constraint reads
\begin{equation}
{\cal C}_{\rm sad}=\frac{1}{R_\star^3}\lambda_1 \lambda_2 \lambda_3 \Theta_{\rm H}(\lambda_1)\Theta_{\rm H}(-\lambda_2)\delta_{\rm D}(\nu_i)\,,
\end{equation}
where the Dirac delta function ensures the gradient to be zero, the Heaviside Theta functions impose the sign of the eigenvalues, the Jacobian $\lambda_1 \lambda_2 \lambda_3=\det \nu_{ij}$ accounts for the volume associated with a saddle point and $R_{\star}={\sigma_2}/{\sigma_1}$.

To predict the mean density map around a saddle point, one has to consider the joint statistics of $(\nu,\nu_i,\nu_{ij})$ together with the density field $\nu'$ at a distance $\bf r$ from the saddle point.
In addition,  the symmetry along the axis of the filament ($i=1$ here) will be broken by imposing the first axis to be oriented in the opposite direction from that of the gradient of the gravitational potential (i.e towards the deepest potential well, the most attractive node). One therefore also has to  consider $\Phi_1$ the derivative of the gravitational potential along the first direction rescaled by its corresponding variance $\sigma_{-1}=\sqrt{\left\langle (\nabla \Phi)^2\right\rangle}$.
Let us gather those 12 fields in a vector $\mathbf{X}=\{\nu',\nu,\nu_1,\nu_2,\nu_3,\nu_{11},\nu_{22},\nu_{33},\nu_{12},\nu_{13},\nu_{23},\Phi_1\}$ whose PDF can be written

\begin{equation}
{\cal P}(\mathbf{X})=\! \frac{1}{\sqrt{{\rm det}| 2\pi \mathbf{C}|}  }
\exp\left(\!-\frac{1}{2}X^{\rm T}
\cdot \mathbf{C}^{-1} \cdot X\right)\,,
\end{equation}
where the covariance matrix $C= \langle  \mathbf{X}\cdot \mathbf{X}^{\rm T} \rangle$ depends on the separation vector $\bf r$ and the linear power spectrum
$P_k(k) $ which can include a filter function on a given scale. In this work,   a  $\Lambda$-CDM power spectrum is used (using the same values for the cosmological parameters as \hagn) with a Gaussian filter defined in Fourier space by
\begin{equation}
    W_{\rm G}(\mathbf{k},L)=
    \frac{1}{(2\pi)^{3/2}}\exp\left(\frac{-k^{2}L^{2}}{2}\right)
    \,,
    \end{equation}
with $L=0.8$Mpc$/h$.
One-point covariances do not depend on the separation but may depend on the spectral parameter $\gamma={\sigma_1^2}/({\sigma_0\sigma_2})$.
The variance of the density field is one by definition, $C_{11}=1$, while the diagonal block corresponding to the saddle position, ${\mathbf C}_0=(C_{ij})_{i,j>1}$, reads
\setlength{\arraycolsep}{1.5pt}
{\footnotesize
 \begin{equation}
\mathbf{C}_{0}\!=\!
\!\! \left(\!\!\!
\begin{array}{ccccccccccc}
1\!&\!0\!&\!0\!&\!0\!&\!-\!\gamma/3\!&\!-\!\gamma/3\!&\!-\!\gamma/3\!&\!0\!&\!0\!&\!0\!&\!0\!\\
0\!&\!1/3&0&0&0&0&0&0&0&0&\!-\!\beta/3\!\\
0\!&\!0&1/3&0&0&0&0&0&0&0&0\!\\
0\!&\!0&0&1/3&0&0&0&0&0&0&0\!\\
-\gamma/3\!&\!0&0&0&1/5&1/15&1/15&0&0&0&0\!\\
-\gamma/3\!&\!0&0&0&1/15&1/5&1/15&0&0&0&0\!\\
-\gamma/3\!&\!0&0&0&1/15&1/15&1/5&0&0&0&0\!\\
0\!&\!0&0&0&0&0&0&1/15&0&0&0\!\\
0\!&\!0&0&0&0&0&0&0&1/15&0&0\!\\
0\!&\!0&0&0&0&0&0&0&0&1/15&0\!\\
0\!&\!-\beta/3&0&0&0&0&0&0&0&0&1/3\!\\
\end{array}
\!\!\!\right)\!, \nonumber
 \end{equation}
 }
with $\beta=\sigma_{0}^{2}/\sigma_{-1}\sigma_{1}$.
The cross correlations between $\nu'$ and the fields at the position of the saddle are to be computed carefully as they depend on both the separation and the orientation of the separation vector in the frame of the Hessian described by the coordinates  with indices $i=1,2,3$. They are explicit function of the shape of the power spectrum and are therefore computed numerically (the angle dependence is analytical, hence only the integration w.r.t $k=|\bf k|$ requires a numerical integration). They read for $j$ between 2 and 12
 \begin{equation}
\hskip -0.1cm   \left\langle \nu' X_j\right\rangle =\frac{\displaystyle\int {\rm d}^3 {\bf k} P_k (k) \prod_{i=1}^3 (-\imath k_i)^{\alpha_i}(\imath k)^{-2 p} \exp \left(\imath {\bf k}\cdot {\bf r}\right)}{\displaystyle\int {\rm d}^3 {\bf k} P_k (k) \int {\rm d}^3 {\bf k} P_k (k) \prod_{i=1}^3 k_i^{2\alpha_i}}\,,
 \end{equation}
 where $p=1$ only for $j=12$ (because of Poisson equation) and zero elsewhere and $\alpha_i$ counts the number of derivatives w.r.t index $i$.
Note that the mean density map around a saddle point of fixed height and curvatures with no symmetry breaking (i.e  not imposing $\Phi_1<0$) is analytical and given by \citep{Codis2015a}
 \begin{multline}
\left\langle \nu'|{\cal S}\right\rangle=
\frac{(\lambda_{1}\!+\!\lambda_2\!+\!\lambda_3) (\left\langle\nu' {\rm tr}\,\nu_{ij}\right\rangle+\gamma \left\langle\nu'\nu\right\rangle )}{1-\gamma ^2}\\
+
\frac{\nu  (\left\langle\nu'\nu\right\rangle+\gamma \left\langle\nu' {\rm tr}\,\nu_{ij}\right\rangle)}{1-\gamma ^2}
+\frac {45}{4}\!\left(\mathbf{\hat r}^{\rm T} \cdot \overline{\mathbf{ H}}\cdot\mathbf{\hat r}\right)\! \left\langle \nu' \!\left(\mathbf{\hat r}^{\rm T} \cdot \overline{\mathbf{ H}}\cdot\mathbf{\hat r}\right)\!\right\rangle,
\nonumber\end{multline}
where $\overline{\mathbf{ H}}$ is the {\sl detraced} Hessian of the density and
$\mathbf{\hat r}={\mathbf{r} }/{r}$.
However, here the goal is to compute this mean map around an arbitrary saddle (marginalising over its height and curvatures) and with symmetry breaking. To do so, a Monte Carlo technique is implemented to compute the integrals of typically 6 dimensions with {\sl Mathematica}.

The mean map marginalised over the direction perpendicular to the filament is shown on the left panel of Figure~\ref{fig:halo-mass-comparison}.
As expected, a filamentary ridge is predicted along the $\lambda_1$ direction with two nodes at about 3 smoothing lengths from the saddle. In the direction perpendicular to the filament, two voids are typically found on both sides of the saddle.
In addition, Figure~\ref{fig:xy-plane-GRF}   also shows  the mean density in a plane perpendicular to the filament and containing the saddle point. As expected the filament cross section is squashed in the direction of the wall ($\lambda_2$).
This squashing will depend on the peak height and therefore on the mass of galaxies and haloes, namely the rarer objects will display a more spherical cross-section and vice versa.
Note that for both plots, 10 millions draws of the fields per point are drawn from  a Gaussian distribution conditioned to having $\nu_i=\nu_{12}=\nu_{13}=\nu_{23}=0$.
All configurations with positive $\Phi_1$ and wrong signs of the eigenvalues are thrown  before computing the mean density $\nu'$ in those configurations with weights $\lambda_1\lambda_2\lambda_3$ (because of the $\nu_i$ condition) times $(\lambda_{1}-\lambda_{2})(\lambda_{2}-\lambda_{3})(\lambda_{1}-\lambda_{3})$ (because of the $\nu_{12}=\nu_{13}=\nu_{23}$ condition).

\subsection{Cosmic evolution of the dark matter maps} \label{sec:cubic}

The above formalism is valid in the Gaussian initial conditions and can in principle be extended perturbatively to the subsequent weakly non-linear cosmic evolution.
For the sake of simplicity, only the mean non-linear evolution of the density distribution around a saddle point of {\sl fixed geometry} is described.
Using a Gram-Charlier expansion \citep{Gay2011} for the joint distribution of the field and its derivative, the first non-Gaussian correction to the mean density map is found to be

\begin{equation}
\langle \delta(\mathbf{r}|{\cal S}) \rangle \sim \langle \delta(\mathbf r|{\cal S}) \rangle_{\rm G}+
\sigma_{0} \left[ \sum_{ijk\le 11}  S_{ijk} H_{ijk}(\mathbf{r})
\right] \,,
\label{eq:NGsaddle}
\end{equation}
where higher order terms ${\cal O}(\sigma^2)$ are neglected.
In Equation~\eqref{eq:NGsaddle}, the $S_{ijk}\equiv \langle X_i X_j X_k  \rangle/\sigma$ coefficients generalise the so-called $S_3\!\equiv\! \langle \delta^3  \rangle/ \langle \delta^2  \rangle^2$ to  expectations of cubic  combinations of the field $\nu\!=\!\delta/\sigma_{0}$
and the components of its gradients, $\nu_{k}$, and its Hessian, $\nu_{ij}$ (rescaled by their respective variance)
evaluated at the running point $\mathbf{r}$ and at the saddle.
In equation~\eqref{eq:NGsaddle}, the  function $H_{ijk}(\mathbf{r})$ only involves known combinations
of the Gaussian covariance matrix $C_{ij}$
evaluated at separation $\mathbf{r}$ (so  $H_{ijk}$ is independent of redshift).
Note importantly that at tree order, the $S_{ijk}$ also do not depend on $\sigma_{0}$, so that the only (degenerate)
dependence on cosmic time $\tau$ and smoothing scale $L$ (over which the saddle is defined) is through $\sigma_{0}(L,\tau)$ in front of the square bracket of
equation~\eqref{eq:NGsaddle}.
For the purpose of this paper, this equation therefore implies that gravitational clustering
will distort and enhance the contours of dark matter density within the frame of the saddle, with a scaling
proportional to  $\sigma_{0}$\footnote{As such, the thickening of filaments provides us with a cosmological probe, though admittedly it might not be the most straightforward one!}. Here the considered scale $L$ can also be related to the typical mass, $M_\star$ of
the population considered so that the local clock becomes $\sigma(M_\star,$redshift).
Hence equation~\eqref{eq:NGsaddle} simply predicts the observed mass and redshift scalings
of the main text.
 In practice,  computing the whole $S_{ijk}$ suite takes us beyond the scope of this
paper and will be investigated elsewhere.

 Notwithstanding, as a first approximation, most of the effect is simply due to the density boost $\nu=\nu_{\cal S}$ at the location of the saddle.
 The corresponding non-Gaussian correction is  simply given by $\sigma_{0}$ multiplied by
 \begin{equation}
\frac{H_2(\nu_{\cal S})}{2}\frac{\xi(r)}{\sigma_{0}^2} \left( C_{12}(r)-  S_{3}\right)
   \,,
\label{eq:NGsaddleLS}
\end{equation}
with $H_2(x)=x^2-1$ the second Hermite polynomial, $C_{12}=\left \langle \rho^2({\bf x})\rho(\bf x+r)\right \rangle_{c}/\sigma_{0}^2\xi$ and again $S_{3}= \left \langle \rho^3\right \rangle_{c}/\sigma_{0}^4$.
Note that at tree order in perturbation theory, in the large separation limit, $ C_{12}(r)-  S _{3}\rightarrow-34/21$. For a saddle point one sigma above the mean, $H_2(\nu_{\cal S})>0$, which means that the non-linear evolution tend to sharpen the density profile around the saddles   (given that the height of the saddle, $\nu_{\cal S}$ is fixed here),
as one would have expected.

Alternatively, excursion set theory \citep{Musso2018} allows us to
predict the typical mass distribution in the vicinity of a given
saddle point (with fixed geometry) and as was done in that paper, the
predicted profile can be displaced via a so-called Zel'dovich boost.
This is shown in Figure~\ref{fig:zeldoboost}, which corresponds to a
cross section through Figure~13 of \cite{Musso2018} where the length have
been rescaled by a factor $\alpha$ and the masses by a factor $\alpha^3$ to
match the smoothing scale used in this paper and to account for
differences arising from the use of a different filter (Gaussian
vs. Top-Hat). Using the same approach, it is also possible to compute
the expected accretion rate of the dark matter halo. One then recovers Figure~12 of
\cite{Musso2018} that is showing that the effect of the saddle point
on the accretion rate decreases as the mass of the halo decreases. In
the (simplistic) picture where dark matter accretion rate correlates with fresh
gas accretion and specific star formation, one then qualitatively recovers the
results of Figure~\ref{fig:3D_ssfr_long}, where the effect of the
cosmic web onto the sSFR decreases with the stellar mass. Indeed, as
these two quantities (dark matter accretion rate and sSFR) only probe the
recent accretion history of the halo, they are sensitive to {\em
differential} effects induced by the saddle point which vanish at
scales much smaller than that of the filament.

In order to compare this Lagrangian prediction to simulations,  the mean total matter distribution was measured around saddles in the \hagn simulation. The low-redshift measurement is shown on the left panel of Figure~\ref{fig:mass-evolution}.
Interestingly, the prediction for Gaussian random fields recovers the qualitative picture found in the \hagn simulation in terms of the geometry of the contours. As expected, the non-linear evolution (not captured by the Gaussian prediction) further contracts the filaments which become more concentrated. As one goes to higher redshifts (right panel of Figure~\ref{fig:mass-evolution}), the contours clearly become closer to the Gaussian prediction.

\section{Kinetic bulk flow near saddle }
\label{sec:vel}
\begin{figure*}
\centering
\includegraphics[width=\textwidth]{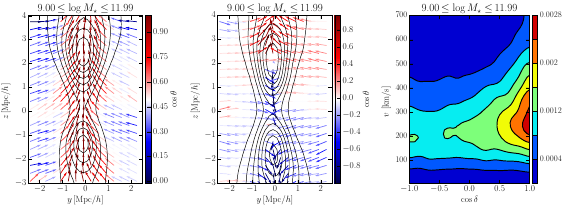}
\caption{{\sl Left} and {\sl middle panels}: Velocity vector of galaxies in the frame of the saddle for masses in the range $10^9$ to $10^{12} M_\odot$ at redshift zero, in the longitudinal plane of the saddle, with the upward direction defined to be towards the node of  highest density ({\sl left}), and without imposing any condition on the direction of the filament at the saddle ({\sl middle}). The arrows represent the unitary velocity vector in the $y-z$ plane, colour-coded by the cosine  of the angle between this vector and the $z-$axis. Note that as the modulus of the velocity near the saddle point is low, the direction of the velocity vector is not very well constrained there. Note also that the structure of the velocity field in the saddle frame is intrinsically complex due to the relative velocity of a saddle w.r.t. the nodes.
The grey contours represent the galaxy number counts.
{\sl Right panel}: Probability distribution function of the velocity field of galaxies within the cosmic web as a function of its modulus $v$, and  the cosine  of its angle $\delta$ with the closest filament. The excess of galaxies with $\cos \delta$ close to unity shows that the bulk of the population appears to be flowing along the filaments in the direction of nodes, i.e.  the high density regions.
}
\label{fig:3D_curv_761_flow}
\end{figure*}

Extending the result of  \cite{Sousbie2008} (which focussed on dark matter), 
let us quantify the geometry of the bulk galactic
velocity flow in the frame of the saddle.
Figure~\ref{fig:3D_curv_761_flow} displays the (normalized) velocity field  of 
galaxies in the frame of the saddle while tracking (left panel) or not (middle panel) the orientation of the saddle, and the PDF of the velocity's modulus and orientation (right panel). 
The velocities of the left and middle panels are computed as previously, i.e. as an average velocity for all galaxies in given 2D bin
and smoothed over 0.3 Mpc/$h$. Note that no flipping w.r.t. the $z$-axis is applied here.
As expected, when the frame is orientated towards the larger node (left panel), the net flow is directed towards 
that node throughout that frame.  
Interestingly, note that the flow actually overshoots the peak of density in that frame, which is in fact 
expected, in so far that the velocity should, at the level of the Zel'dovich approximation, point 
towards the minimum of the potential, whose peak is typically further away from the saddle.
When the orientation of the saddle is ignored (middle panel), one recovers a `saddle-like' geometry for the flow,
i.e. the saddle point locally repels the flow longitudinally but attracts it transversally. 
The right panel  is consistent with  Figure~6 of \cite{Sousbie2008}, but applies now to galaxies 
in \hagn. The  PDF velocity orientation and moduli present a  tail of high velocities (at $\cos \delta < 0$), corresponding to galaxies converging transversally  towards filaments.

The geometry of the flow displayed in Figure~\ref{fig:3D_curv_761_flow}, together with the distinct initial population distribution (and accretion history) for the progenitor  of high and low mass galaxies 
allow us to understand their cosmic evolution presented in the main text. 
On top of this passive  advection, Section~\ref{sec:hidden_variable} argues that the tides of the saddle may impact directly 
dark halo growth while shifting the conditional mean and  co-variances of the accretion rate,
and galactic \vsig or \ssfr  while biasing spin (hence cold gas) acquisition.

\section{Statistical origin of residuals}
\label{sec:stat}

Let us finally discuss the statistical basis of the procedure described in the main text to study second order effects beyond the mass and density and capture the origin of these hidden variables. 
When attempting to disentangle the  specific role of tides from that of the local density 
and/or that of the dark halo mass, we are facing a statistical mediation problem \citep[see e.g.][]{Wright1934,Sobel1982,BaronKenny1986}, 
in that we aim to determine if the tidal tensor plays a specific role impacting the \ssfr (or \vsig or age etc.) which is not already encoded in other quantities such as density and dark matter mass
(which also vary away from the saddle, but typically with  different maps in that frame).
For the sake of being concrete, let us assume that the effect of the tides can be 
summed up by a scalar field $\alpha$ (e.g. the squared sum of the difference of the eigenvalues of the Tidal tensor,
$\alpha =\sum (\lambda_i-\lambda_j)^2$, which quantifies the anisotropy of the collapse,
or the net flux of advected angular momentum, etc.). Our purpose is to extract the map $\alpha(\mathbf{r})$ and check  its structure relative to the saddle.

\subsection{Conditional mediation}

Let us motivate the  procedure used in Section~\ref{sec:hidden_variable} while relying on a statistical description of the random variables describing the various fields at some given position away from 
the saddle.
Let us first assume for simplicity that the field
  $\mathbf{X}=(\ssfr,\delta \equiv\log\rho,m \equiv \log M_{\rm DH},\alpha)$ 
  obeys  a centred\footnote{the PDF is assumed to be centred on the 
  mean value of the field averaged over the whole map}  joint  Gaussian statistics:
\begin{equation}
{\rm PDF}(\ssfr, \delta, m,\alpha)\!=\! \frac{1}{\!\sqrt{{\rm det}({\mathbf C_0})}} \exp\left(-\frac{1}{2} \mathbf{X}^{T}\cdot {\mathbf C_0}^{-1}\cdot \mathbf{X}\right)\notag\,,
\end{equation}
where $\mathbf C_0$ is the matrix of the covariance of the four fields, which we will also assume for now to be position independent (but see below). 
Note the change of variable to $m$ and $\delta$ which are likely to behave more like 
Gaussian variables than $M_{\rm DM}$ and $\rho$.

Applying Bayes' theorem, we can compute the conditional PDF$(\ssfr|\delta,m,\alpha)= {\rm PDF}(\ssfr,\delta,m,\alpha)/{\rm PDF}(\delta,m,\alpha) $,
where PDF$(\delta,m,\alpha)$ is the marginal (after integration over \ssfr). 
From this conditional PDF
 the expectation  $\ssfr(\mathbf{r})\equiv\langle \ssfr| \delta,m, \alpha,\mathbf{r} \rangle $
subject to the constraint of the three fields $\delta(\mathbf{r}),m(\mathbf{r})$ and $\alpha(\mathbf{r})$
 reads
\begin{align}
 \ssfr(\mathbf{r})
  =&
 \left(\! \begin{array}{ccc}
\langle \ssfr\, \delta\rangle, & \langle \ssfr\, m \rangle,
& \langle \ssfr\, \alpha \rangle \end{array}\! \right) \label{eq:cond}
\\ 
&
\cdot\left(\begin{array}{ccc}
 \langle \delta^2\rangle & \langle \delta m\rangle &  \langle \delta \alpha\rangle\\
\langle  \delta m \rangle &  \langle m^2\rangle & \langle \alpha m \rangle\\
\langle  \delta \alpha \rangle  & \langle  \alpha  m\rangle & \langle \alpha^2\rangle
\end{array}\right)^{-1}\!\! \cdot\left(\begin{array}{c}
\delta(\mathbf{r})\\
m(\mathbf{r})\\
\alpha(\mathbf{r})\\
\end{array}\right) \equiv\sum_{i>1} \beta_i X_i ,\notag
\end{align}
so that
 the conditional  $\ssfr(\mathbf{r})$  is simply a linear combination of the three $X_i$ maps, $\delta(\mathbf{r})$, $m(\mathbf{r})$
and $\alpha(\mathbf{r})$ (with coefficients $\beta_i$ involving the  covariances)\footnote{This relationship could have also been obtained by principal component analysis in the extended $\mathbf{X}$ space: it would have led to the same sets of covariances as linear coefficients.}. 
Let us now take the statistical expectation of this equation {\sl at a given pixel}.
Subtracting the contribution of $\langle \delta(\mathbf{r})
\rangle$ and $ \langle m(\mathbf{r}) \rangle$  from the measured $\langle\ssfr(\mathbf{r})\rangle$ 
(while using a linear fit to the simulation to estimate the $\beta_i$ since we do not know a priori 
what the covariances involving $\alpha$ might be\footnote{Note that for an explicit choice of $\alpha$ we could have extracted the covariances entering 
equation~\eqref{eq:cond} from the simulation and estimated the  $\beta_i$ accordingly.})
and focusing on residuals  provides a position-dependent estimate of the field $\langle \alpha(\mathbf{r})\rangle$. 
If its amplitude is statistically significant, its geometry may tell us 
if it is  consistent with the nature of the mediating physical process, as discussed in the main text.

If we relaxed the assumption of Gaussian statistics, the conditionals derived from  an Gram-Charlier expansion of the joint PDF \citep{Gay2011}  would lead (to leading order in non Gaussianity) to the  mapping
\begin{equation}
 \ssfr(\mathbf{r})=\sum_{i>1} \beta_i X_i + \sum_{i,j>1} \beta_{ij}  X_i X_j +\cdots
\end{equation}
 where 
$\beta_i$  (resp. $\beta_{ij}$) are   functions of the  second resp. second and third order cumulants of the fields
(such as $\langle \delta^2 \alpha \rangle$ etc). 
Once again we could subtract the (up to quadratic) fitted contribution of  $\langle\delta(\mathbf{r})\rangle$ and $\langle m(\mathbf{r})\rangle$  from the measured $\langle \ssfr(\mathbf{r})\rangle$ so as to fit the  manifold
of the $\mathbf{X}$ samples. Unfortunately,
in this non-linear regime, 
the expectations would not compute any more, $\langle X_i  X_j\rangle \neq \langle X_i \rangle \langle X_j\rangle$  
and the residuals will also involve terms such as $\langle \delta \alpha \rangle$,
$\langle m \alpha \rangle$ or $\langle  \alpha^2 \rangle$.

 In practice though, the extracted relationships 
 expressed in terms of  $\delta =\log \rho$ and $m=\log M_{\rm DH}$ do in fact look fairly linear, 
 see e.g. the top panels of Figure~\ref{fig:3D_curv_761_delta},
 which favours the assumption of Gaussianity, as was assumed in the main text. We also checked that  relationships such as \eqref{eq:cond} did not 
significantly vary with position within the saddle frame (by marginalising over sub regions within the frame).   
Finally we used the median to extract the $\beta_i$ coefficients, as it is a more robust estimator.

As a word of caution, it should nevertheless  be stressed that 
since we are aiming to extract a secondary effect (beyond mass and density), the impact of departure from  our 
assumptions may prove to 
be of the same order as the sought signal. Eventually,  larger statistical samples  may
 allow us to statistically  disentangle more robustly the various processes.
Note finally  that carrying out the analysis at fixed {\sl stellar} mass allows us to avoid  the bimodality of some physical parameters which would clearly have broken the assumption of 
joint Gaussian statistics.

\subsection{Mediation of multiple causes}

An alternative strategy  to address the fact that more than one variable impact \vsig (and/or \ssfr, age etc.) 
is to sample over narrow bins of stellar and dark halo mass, local density and position $\mathbf{r}$ 
within the frame of the cosmic web, and estimate the full joint PDF. This is challenging for a sample of only $10^5$ galaxies, hence can only be applied to relatively large bins in practice. 
We attempted to disentangle halo mass, density and tidal effects by computing the residuals of \vsig (and/or \ssfr, age) from median halo mass mapping in a given stellar mass and density bin. 
We found comparable residuals to those shown in Figure~\ref{fig:3D_curv_761_delta}. 
However, given the size of the bins we use, we cannot draw any definitive conclusions here. Simulations with more statistics should be able to address this difficult point in the future.

\end{document}